\documentclass[preprint,12pt]{elsarticle}

\usepackage{amssymb}
\usepackage{amsmath}

\newtheorem{prop}{Proposition}

\newtheorem{defn}{Definition}

\usepackage{pmat}

\journal{Information Sciences}
\begin{document}

\begin{frontmatter}


\title{Chernoff Information between  Gaussian Trees}

\author[Tsing]{Binglin Li}
\ead{libl13@mails.tsinghua.edu.cn}
\cortext[mycorrespondingauthor]{Corresponding author}
\author[LSU]{Shuangqing Wei\corref{mycorrespondingauthor}}
\ead{swei@lsu.edu}
\author[Tsing]{Yue Wang}
\ead{wangyue@mail.tsinghua.edu.cn}
\author[Tsing]{Jian Yuan}
\ead{jyuan@mail.tsinghua.edu.cn}
\address[Tsing]{Department of Electronic Engineering, Tsinghua University, Beijing, P. R. China, 100084}
\address[LSU]{School of Electrical Engineering and Computer Science, Louisiana State University, Baton Rouge, LA 70803, USA}

\begin{abstract}
In this paper, we aim to provide a systematic study of the relationship between Chernoff information and topological, as well as algebraic properties of the corresponding Gaussian tree graphs for the underlying graphical testing problems. We first show the relationship between Chernoff information and generalized eigenvalues of the associated covariance matrices. It is then proved that Chernoff information between two Gaussian trees sharing certain local subtree structures can be transformed into that of two smaller trees. Under our proposed grafting operations, bottleneck Gaussian trees, namely, Gaussian trees connected by one such operation, can thus be simplified into two 3-node Gaussian trees, whose topologies and edge weights are subject to the specifics of the operation.  Thereafter, we provide a thorough study about how Chernoff information changes when small differences are accumulated into bigger ones via concatenated  grafting operations. It is shown that the two Gaussian trees connected by more than one grafting operation may not have bigger Chernoff information than that of one grafting operation unless these grafting operations are separate and independent. At the end, we propose an optimal linear dimensional reduction method related to generalized eigenvalues.
\end{abstract}


\end{frontmatter}
\section{Introduction}

 Gaussian graphical models\cite{GOMEZVILLEGAS2014115,XIU2018187,LARRANAGA2013109} have great successes in representing the dependence of multiple Gaussian random variables in the fields of economics\cite{dobra2010modeling}, social networks\cite{vega2007complex,wasserman1994social}, biology\cite{durbin1998biological,ahmed2008time}, image recognition\cite{geman1984stochastic,besag1986statistical}  and so on. Among Gaussian graphical models, we are particularly interested in Gaussian trees due to its sparse structure as well as existing computationally efficient algorithms in learning the underlying topologies.

\subsection{Motivation and related works}

The error events in learning problems have two different types, namely, conditional error events and average error events. Conditional error events mean the error events between true model and the learned one when we learn an  approximate model from the output data\cite{Tan.TSP.Ana.2010,tan2011large,jog2015model}. This kind of errors also
appear in network\cite{qiao1995,qiao2000}.
We can use Kullback-Leibler (KL) distance to quantify this error exponent on distinguishing true model from the learnt one\cite{kullback1951information}.

Our study belongs to the other type of error events, namely, average error events. We consider error exponents of $M$-ary hypothesis testing against a set of Gaussian trees. That is to say, we want to discern $M$ distributions based on a sequence of data drawn independently from one of these distributions.
Chernoff information is the proper metric in such testing problems, as compared with KL distances\cite{chernoff1952measure,westover2008asymptotic}. Chernoff information is very complex to be calculated, so researchers often approximate it or calculate a bound\cite{santhanam2012information,sason2015tight,sason2015tight1}.

Instead of considering an arbitrary set of Gaussian tree graphs, we are mainly interested in quantifying the relationship between Chernoff information and  variation of  graph structures built from local differences through a sequence of graphical operations defined by us. In literature, there are existing works in regards to how local variation on graphical structure affects KL distances or symmetrize KL distances \cite{Tan.TSP.Ana.2010,tan2011large,jog2015model}, due to  presumed difficulties in evaluating Chernoff information. Our work, to the best of our knowledge, is the first one investigating such relationship from Chernoff information point of view.

In addition, our work can be deemed as  sensitivity study of how graph structure  differences affect information theoretical metrics\cite{GOMEZVILLEGAS2014115,GOMEZVILLEGAS201112409}.
When the topological differences between Gaussian trees are large, the classification problem is relatively easy.  We deal with classification between Gaussian graphs with minor local topological differences and show how these differences affect the distinguishing ability between Gaussian trees.

\begin{figure}
        \centering
        \includegraphics[width=8cm]{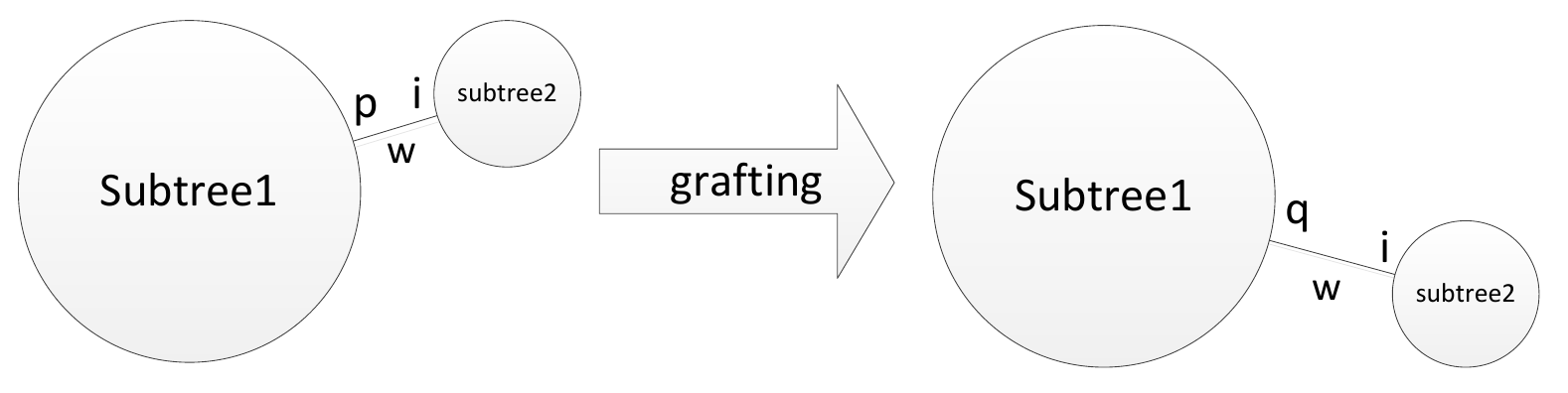}\\
        \caption{$G_2$ is obtained from $G_1$ by grafting operation}\label{grafting}
    \end{figure}

The error exponent of an $M$-ary hypothesis testing problem is the minimum pair-wise Chernoff information among them\cite{westover2008asymptotic}. So we focus on pair-wise topological differences at first. We come up with a typical topological pair, namely two trees connected by one grafting operation, which is deemed to represent the case where two Gaussian trees have a minimal topological difference. Grafting is a kind of topological operation by cutting down a subtree from another tree and pasting it to another location as shown in Fig.~\ref{grafting}. In this figure, $i,p,q$ are the nodes in both trees, representing random variables $X_i,X_p,X_q$. We separate subtree1 and subtree2 by cutting the edge $e_{ip}$ with weight $w$ and paste subtree2 to subtree1 by adding new edge $e_{iq}$ with the same weight $w$. We call $i,p,q$ anchor nodes of this grafting operation.
Our formulation is for the purpose of identifying the contributing factors to Chernoff
information in distinguishing Gaussian trees with minimum topological
difference, which can be deemed as a worst case scenario.

In this paper, we are  particularly interested in finding out the determining factors of Chernoff information in distinguishing Gaussian trees with the
same amount of randomness, i.e. sharing the same determinant of their
covariance matrices and normalized variances.
The assumptions are for the ease of analysis and providing insights, as shown later in our results.
Generalized eigenvalues of covariance matrices $\Sigma_1$ and $\Sigma_2$ are often considered when judging the difference between two Gaussian graphs\cite{5540052,6247965}. We tie the Chernoff information to generalized eigenvalues, which are further shown to be determined by local structure and topological differences between two graphs.

Furthermore, we want to learn how Chernoff information changes when small differences evolve into big ones. That is to say, we want to analyze Chernoff information of two Gaussian trees connected by more than one grafting operation. Our initial expectation was that bigger difference in topology between two Gaussian trees leads to larger Chernoff information. But this may not be true for all situations. We start from a simple case, namely, two successive grafting operations, to study how Chernoff information changes when small differences are added up.

In practical scenarios, we may not have access to all the output of the model. Instead, we may have some constraint on observation costs, which prompts us to reduce the dimension of observation vectors  in order to meet such  constraints\cite{NOWAKOWSKA201674,GUAN2013147}. A good choice here is doing linear dimension reduction in collection stage. Former study deals with classification among multiple Gaussian trees, but we only deal with a $2$-ary hypothesis testing with dimensional reduction.

Traditional dimension reduction methods, such as Principal Component Analysis (PCA) and other Representation Learning\cite{6472238}, aim to find the optimal features with maximum information. They only consider a distribution at one time and don't care how we use the resulting low-dimensional data. But our method considers two hypothesis at the same time and want to keep the most information for classification. Some features may be important in PCA, but invalid in our method because these features in two hypothesis are similar. Our objective function here is the Chernoff information between the resulting low-dimensional distributions. We linearly transform an $N$ dimensional Gaussian vector $x$ to an $N_O < N$ dimensional vector $\mathbf{y}=\mathbf{A}\mathbf{x}$, through an $N_O \times N$ matrix $\mathbf{A}$. We provide an optimal linear transformation $\mathbf{A}$ which can maximize Chernoff information in the classification of two hypothesis.

\subsection{Contribution}

The primary contribution of this paper is to show the relationship among tree topology, generalized eigenvalues of covariance matrices and Chernoff information. In particular, we show
how tree topological changes are represented in algebraic expression and how tree topological changes affect the Chernoff information between two trees.

Our major and novel results can be summarized as follows.
First of all, we provide the relationship between Chernoff information and covariance matrices $\Sigma_1$ and $\Sigma_2$'s generalized eigenvalues. We show that Chernoff information between two Gaussian trees is a function of  generalized eigenvalues and in particular an important parameter $\lambda^*$, which can be found in an equation of generalized eigenvalues. Then we show two special operations, namely, adding operation and division operation, which can keep Chernoff information between two Gaussian trees unchanged. These results can tell us how to reduce the
complexity of computing Chernoff information between two
Gaussian trees sharing the same local parameters.

We also show that Gaussian trees connected by a grafting operation are the bottleneck under the assumption of the same entropy and normalized variance.
We further prove that Chernoff information between
two Gaussian trees connected by a
grafting operation is the same as that of two $3$-node
Gaussian trees, whose topologies and edge weights are subject
to the underlying graph operation. Grafting operation is for the purpose of constructing trees of minimal topological difference.  Chernoff information between trees connected by a grafting operation can be calculated exactly.

As Gaussian trees we actually deal with  have much more differences, we also study how Chernoff information changes when small bottleneck differences accumulate into large ones. The result shows that Chernoff information may not increase when small differences accumulate, unless these grafting operations are separate from each other. In special grafting chains where grafting operations are decoupled, we can conclude $\lambda^*=0.5$ in every pairs of Gaussian trees and the minimal Chernoff information can only appear in adjacent pairs of the chain.

At last, we consider the dimension reduction problem. We provide an optimal linear dimension reduction and show the relationship between optimal transformation matrix and generalized eigenvalues.

\subsection{Comparison with former paper}

Our early results on this topic are published in a conference paper \cite{binglin}. As a summery, below is a list of the new results presented only in this journal paper as compared with \cite{binglin}.

\noindent$\bullet$ We provide the relationship between Chernoff information and generalized eigenvalues of covariance matrices.

\noindent$\bullet$ In \cite{binglin}, we have already proved that adding operation and division operation
can keep Chernoff information between two Gaussian trees unchanged. Now we connect these operations with the changes of generalized eigenvalues and provide a new proof.

\noindent$\bullet$ In \cite{binglin}, we mainly studied two Gaussian trees connected by one grafting operation and found the relationship between Chernoff information of them and the maximum generalized eigenvalue of $\Sigma_1$ and $\Sigma_2$.
In this paper, we extend our study from one grafting operation case to more grafting operations in order to study how Chernoff information changes when small differences are added up.

\noindent$\bullet$ We define independent grafting operations which can allow us build a partial ordering among the trees in a grafting chain.

\noindent$\bullet$ In this paper, we provide an optimal and linear dimension reduction method based on generalized eigenvalues for an arbitrary $N_O \geq 1$, the size of lower-dimension vector, rather than just $N_O=1$ as in \cite{binglin}. Such generalization is non-trivial as exhibited in its proof.

\subsection{Organization}
The paper is organized as follows. Section \ref{section2} presents the formulation and system model of this paper. Section \ref{section3} shows the relationship between Chernoff information and generalized eigenvalues.  The definitions of grafting operation and Chernoff information of bottleneck Gaussian trees are provided in Section \ref{section4}. A classification of two successive grafting operation is shown in Section \ref{section5} and a partial ordering comparison result of grafting chains is presented in Section \ref{section6}. Results about dimension reduction problem are in Section \ref{section7}. At last,   conclusion is drawn in Section \ref{section8}.

\section{Formulation}\label{section2}

Gaussian tree models can represent the dependence of multiple Gaussian random variables by tree topologies. For simplification, we normalize the variance of all Gaussian variables to be $1$ and the mean values to be $0$. For an $N$-node tree $\mathbf{G} = (V, E, W)$ with  vertex set $V=\{1,\dots,N\}$, edge set $E=\big\{e_{ij}|(i,j)\subset V\times
V\big\}$ and edge weights set $W=\{w_{ij}\in [-1,1]|e_{ij}\in E\}$, $E$ satisfies $|E|=N-1$ and contains no cycles. A vector of Gaussian variables $\mathbf{x}=[x_1,x_2,\dots,x_N]^T\sim N(\mathbf{0},\Sigma)$ is said to be a normalized Gaussian distribution on the tree $\mathbf{G} = (V, E, W)$ if
    \begin{align}
    \sigma_{ij}=
    \begin{cases}
    1&\quad i=j\\
    w_{ij}&\quad e_{ij}\in E\\
    w_{im}w_{mn}\dots w_{pj}&\quad e_{ij}\notin E
    \end{cases}
    \end{align}
    where $\sigma_{ij}$ is the $(i,j)$ term of $\Sigma$ and $e_{im}e_{mn}\dots e_{pj}$ is the unique path from node $i$ to node $j$.

    A normalized covariance matrix of a Gaussian tree has a very simple inverse matrix and determinant, as shown in Proposition \ref{thm1} which has been proved  in \cite{binglin}.

    \begin{prop}\label{thm1}
    Assume $\Sigma$ is a normalized covariance matrix of Gaussian tree  $G=(E,V,W)$, so $|\Sigma|=\prod_{e_{ij}\in E}(1-w_{ij}^2)$ and the elements $[u_{ij}]$ of $\Sigma^{-1}$ follow the following expressions:
    \begin{align}
    u_{ij}=
    \begin{cases}
    \frac{-w_{ij}}{1-w_{ij}^2}&\quad i\neq j~\text{and}~e_{ij}\in E\\
    0&\quad i\neq j~\text{and}~e_{ij}\notin E\\
    1+\sum_{p:e_{ip}\in E}\frac{w_{ip}^2}{1-w_{ip}^2}& \quad i=j.
    \end{cases}
    \end{align}
    \end{prop}

     A set of normalized Gaussian trees, namely, $\mathbf{G}_k=(V, E_k,
W_k)~k=1,2,\dots,M$, have prior probabilities given by $\pi_1,\pi_2,
\dots, \pi_M$, where $M$ is the number of the trees. They share the same entropy, and thus the same determinant
of their covariance matrices $\Sigma_k=[\sigma_{ij}^{(k)}]$. The same entropy means the same amount of randomness.
This assumption can let us compare these Gaussian trees fairly, and focus on the difference of topologies rather than a single variable.
We want to do
an $M$-ary hypothesis testing to find out from which Gaussian tree
the data sequence $\mathbf{X}=[\mathbf{x}_1,\dots,\mathbf{x}_T]$
($\mathbf{x}_l=[x_{1,l},\dots,x_{N,l}]'$) comes from. We define the average error
probability of the hypothesis testing to be $P_e$, and let $E_e = \lim_{ T\rightarrow\infty}\frac{-\ln P_e}{T}$ be
the resulting error exponent,  which
depends on the smallest Chernoff information between the
trees \cite{westover2008asymptotic}, namely,
    \begin{align}
    E_e=\min_{1\leq i\neq j\leq M} CI(\Sigma_i||\Sigma_j)
    \end{align}
    where $CI(\Sigma_i||\Sigma_j)$ is the Chernoff information between the $i^{th}$ and $j^{th}$ trees.

For two $0$-mean and N-dim Gaussian joint distributions, $\mathbf{x}_1\sim N(0,\Sigma_1)$ and $\mathbf{x}_2\sim N(0,\Sigma_2)$,
    \begin{align}
    f_\Sigma(\mathbf{x})=\frac{1}{{(2\pi)}^{N/2}{|\Sigma|}^{1/2}}\exp{\big(-\frac{\mathbf{x}^T\Sigma^{-1}\mathbf{x}}{2}\big)}
\\
    D(\Sigma_1||\Sigma_2)=
    \frac{1}{2}\log\frac{|\Sigma_2|}{|\Sigma_1|}+\frac{1}{2}tr(\Sigma_2^{-1}\Sigma_1)-\frac{N}{2}\label{D}
    \end{align}
    where $tr(\mathbf{X})=\sum_{i}x_{ii}$ is the trace of the matrix.

    We define a new distribution $N(0,\Sigma_\lambda)$ in the exponential family of the $N(0,\Sigma_1)$ and $N(0,\Sigma_2)$, namely
    \begin{align}
    \Sigma_\lambda^{-1}=\Sigma_1^{-1}\lambda+\Sigma_2^{-1}(1-\lambda)
    \end{align}
so that Chernoff information is given as below
    \begin{align}
    CI(\Sigma_1||\Sigma_2)=D(\Sigma_{\lambda^*}||\Sigma_2)=D(\Sigma_{\lambda^*}||\Sigma_1)\label{lambda}
    \end{align}
    where $\lambda^*$ is the unique point in $[0,1]$ at which  the latter equation is satisfied\cite{cover2012elements}.

Given $|\Sigma_1|=|\Sigma_2|$, $tr(\Sigma_1^{-1}\Sigma_{\lambda^*})=tr(\Sigma_2^{-1}\Sigma_{\lambda^*})$.
    And $\lambda^*\ast tr(\Sigma_1^{-1}\Sigma_{\lambda^*})+(1-\lambda^*)\ast tr(\Sigma_2^{-1}\Sigma_{\lambda^*})=tr(\Sigma_{\lambda^*}^{-1}\Sigma_{\lambda^*})=N$. So $tr(\Sigma_1^{-1}\Sigma_{\lambda^*})=tr(\Sigma_2^{-1}\Sigma_{\lambda^*})=N$ and $CI(\Sigma_1||\Sigma_2)=\frac{1}{2}\log\frac{|\Sigma_1|}{|\Sigma_{\lambda^*}|}=\frac{1}{2}\log\frac{|\Sigma_2|}{|\Sigma_{\lambda^*}|}$.

    We already know that the overall Chernoff information in an $M$-ary testing is bottle-necked by the minimum pair-wise difference, thus we next focus on the calculation of Chernoff information of  pair-wise Gaussian trees.

    Further research will show that Chernoff information can be determined by the generalized eigenvalues of covariance matrices
    $\Sigma_1$ and $\Sigma_2$. And $\lambda^*$ is also an important parameter in the calculation of Chernoff information.

In an $M$-ary Hypothesis testing problem, the overall Chernoff information is bottle-necked by the minimum pair-wise difference. It has been shown in \cite{jog2015model} that the KL distance between a true Gaussian tree and a learned one is bottle-necked when the learned tree only differs from the true one by removing a leaf and pasting to anther location. So in the rest of this paper, we start from Gaussian trees connected by grafting operation, which is the topological operation with the minimum difference in Chernoff information metric.

    But many cases are more complex than that. We have to learn more about big difference between two Gaussian trees. For simplification, we consider a direct extension of grafting operation, namely the concatenation of more grafting operations.
    That is to say, we have a chain of Gaussian trees in which every tree can be obtained by its adjacent neighbor via a grafting operation. Our expectation is that we can distinguish two Gaussian trees  more easily when they are closer with each other in the chain.
    The expectation is true for independent grafting operations, not for dependent grafting chains as shown in section \ref{section6}. If the grafting operations in the chain are overlapping in topology, Chernoff information between adjacent trees may be larger than that of farther tree-pairs.

\section{Chernoff information of two Gaussian trees}\label{section3}

    Chernoff information is often used to measure the difficulty when we distinguish two distributions.
    But this parameter is always too complex to be calculated exactly, which restricts its application. At the same time, we have little insights about the relationship between Chernoff information and structure difference between Gaussian graphs. We want to analyze the relationship and find ways to simplify the calculation of Chernoff information in certain cases.

\subsection{Relationship between Chernoff information and generalized eigenvalues}

    Chernoff information and generalized eigenvalues are both important parameters to describe the difference between Gaussian graph models.
    Here we will show the relationship between Chernoff information of two Gaussian trees and  generalized eigenvalues of their covariance matrices $\Sigma_1$ and $\Sigma_2$ under the assumption of same entropy and normalized covariance matrix.

    \begin{prop}\label{prop1}
    For two $N$-node Gaussian distributions who have the same entropy and normalized covariance matrix $\Sigma_1,\Sigma_2$, their Chernoff information satisfies
    \begin{align}
    CI(\Sigma_1||\Sigma_2)=
    \frac{1}{2} \sum_i \left\{\ln\{(1-\lambda^*) \sqrt{\lambda_i}+\frac{\lambda^*}{\sqrt{\lambda_i}}\}\right\}
    \end{align}
    where $\{\lambda_i\}$ are the generalized eigenvalues of $\Sigma_1,\Sigma_2$, namely eigenvalues of $\Sigma_1\Sigma_2^{-1}$,  and
    $\lambda^*$ is the unique root in $[0,1]$ of
    \begin{align}
    \sum_{i} \frac{1}{\lambda^*+(1-\lambda^*)\lambda_i}=\sum_{i} \frac{\lambda_i}{\lambda^*+(1-\lambda^*)\lambda_i}\label{1}
    \end{align}
    \end{prop}

    We can prove this proposition from the definition of Chernoff information, as shown in \ref{A1}.

    We can also show that
    \begin{align}
    (1-\lambda^*)\sum_{i} \frac{\lambda_i}{\lambda^*+(1-\lambda^*)\lambda_i}+&\lambda^*\sum_{i} \frac{1}{\lambda^*+(1-\lambda^*)\lambda_i}=N\label{2}
    \end{align}

    From equations (\ref{1}) and (\ref{2}), we can conclude
    \begin{align}
    \sum_{i} \frac{1}{\lambda^*+(1-\lambda^*)\lambda_i}=\sum_{i} \frac{\lambda_i}{\lambda^*+(1-\lambda^*)\lambda_i}=N
    \end{align}

    We  find that  generalized eigenvalues of covariance matrices $\Sigma_1$ and $\Sigma_2$ are the key parameters of Chernoff information. We can get  Chernoff information with these $N$ generalized eigenvalues, so these $N$ parameters contain all the information about the difference between two Gaussian trees.
    The generalized eigenvalues are so important that we need more property about them.

\begin{figure}
    \centering
    \includegraphics[width=4.5cm]{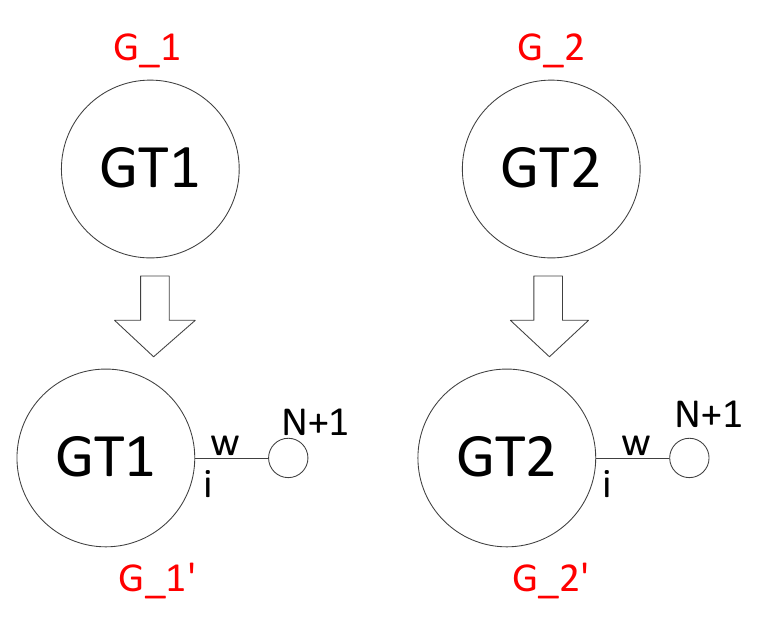}\vline
    \includegraphics[width=8cm]{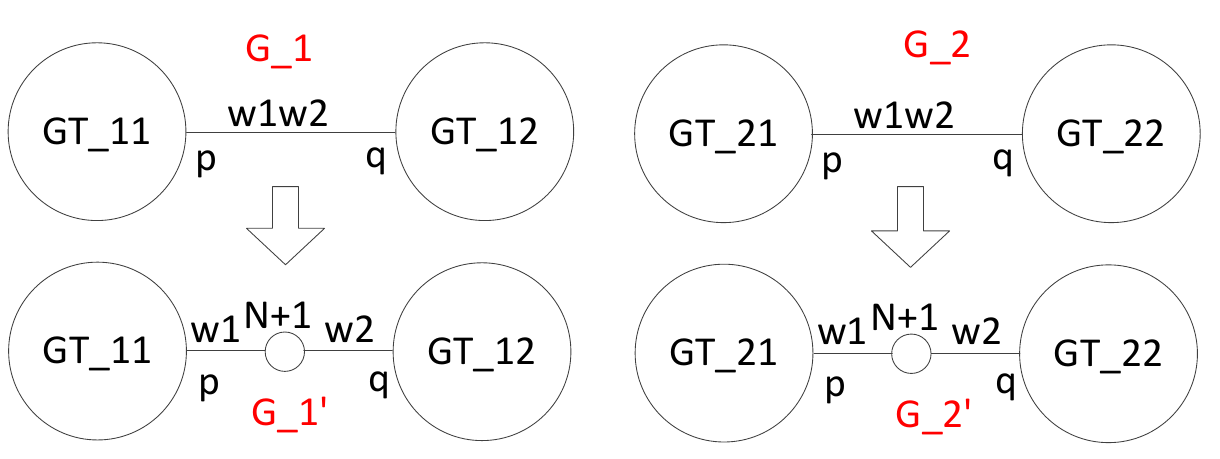}\\
    \caption{Adding and Division operations of two trees}\label{division}\label{adding}
    \end{figure}

    Before we deal with the generalized eigenvalues, we define two special operations on two Gaussian trees, namely, adding operation and division operation. Adding operation and division operation are shown in Fig. \ref{adding}. For adding operation, we add the same leaf node $N+1$, which has the same neighbor $i$ with weight $w$, to both trees. Division operation only appears when two trees have the same edge $e_{pq}$ with the same weight $w_1w_2$, for which we split this edge into two edges and add a node $N+1$ in the path of $p-q$ which has edges $e_{(N+1)p}$ and $e_{(N+1)q}$ with weights $w_1$ and $w_2$, respectively. Their inverse operations are cutting  and merging operations respectively.
     Note that the original trees of division operation should have the same $e_{pq}$ edge and the other parts of the trees can be arbitrary. Similarly, the original trees of cutting operation should have the same leaf $N+1$ while original trees of merging operation should have the same $3$-node path of $(p,N+1,q)$.

    Next we will show how generalized eigenvalues change after adding or division  operation.
    \begin{prop}\label{prop2}
    Assume that Gaussian trees $G_{1}'$ and $G_{2}'$ are obtained from $G_{1}$ and $G_{2}$ by adding operation or division operation. Their covariance matrices are $\Sigma_1',\Sigma_2',\Sigma_1,\Sigma_2$ respectively. The generalized eigenvalues of $(\Sigma_1',\Sigma_2')$ are the same with that of $(\Sigma_1,\Sigma_2)$ except a newly added unit eigenvalue.
    \end{prop}

        The proposition is proved in  \ref{A2}.

    \begin{prop}\label{prop3}
   Assume that Gaussian trees $G_{1}'$ and $G_{2}'$ are obtained from $G_{1}$ and $G_{2}$ by adding operation or division operation. Their covariance matrices are $\Sigma_1',\Sigma_2',\Sigma_1,\Sigma_2$ respectively. The optimal parameter $\lambda^*$ of $(\Sigma_1',\Sigma_2')$ is the same with that of $(\Sigma_1,\Sigma_2)$.
    \end{prop}

    The proposition can be proved by using the former propositions. Former results indicate that $\lambda^*$ is unique in the range of $[0,1]$. So if $\lambda^*$ satisfies equation (\ref{1}), then
    \begin{align}
    &\sum_{i} \frac{1}{\lambda^*+(1-\lambda^*)\lambda_i}+\frac{1}{\lambda^*+(1-\lambda^*)\times1}=\nonumber\\
    &\sum_{i} \frac{\lambda_i}{\lambda^*+(1-\lambda^*)\lambda_i}+\frac{1}{\lambda^*+(1-\lambda^*)\times1}
    \end{align} where $\{\lambda_i\}$ are the generalized eigenvalues of $\Sigma_1,\Sigma_2$.
    The trees after adding or division operation have the same optimal parameter $\lambda^*$ with the original one.

    These propositions indicate that adding operation and division operation are quite special operations when we calculate  Chernoff information. If  trees are connected by adding operation or division operation, we have some simple ways to calculate the Chernoff information.

    \begin{prop}\label{prop4}
    Assuming two $N$-node Gaussian tree models $\mathbf{G}_1=(V,E_1,W_1)$,
    $\mathbf{G}_2=(V,E_2,W_2)$ and two Gaussian trees $\mathbf{G}_1'=(V\cup\{i\}, E_1', W_1'),\mathbf{G}_2'=(V\cup\{i\}, E_2', W_2')$ attained from $\mathbf{G}_1,\mathbf{G}_2$ by adding operation or division operation, $CI(\mathbf{G}_1'||\mathbf{G}_2')= CI(\mathbf{G}_1||\mathbf{G}_2)$.
    \end{prop}

    Proposition \ref{prop4} can be proved by applying Proposition \ref{prop2} and \ref{prop3} in Proposition \ref{prop1}. The two pairs of trees have the same $\lambda^*$ and the same generalized eigenvalues except $1$. So Proposition \ref{prop1} tells us that they have the same Chernoff information.

\begin{figure}
    \centering
    \includegraphics[width=8.5cm]{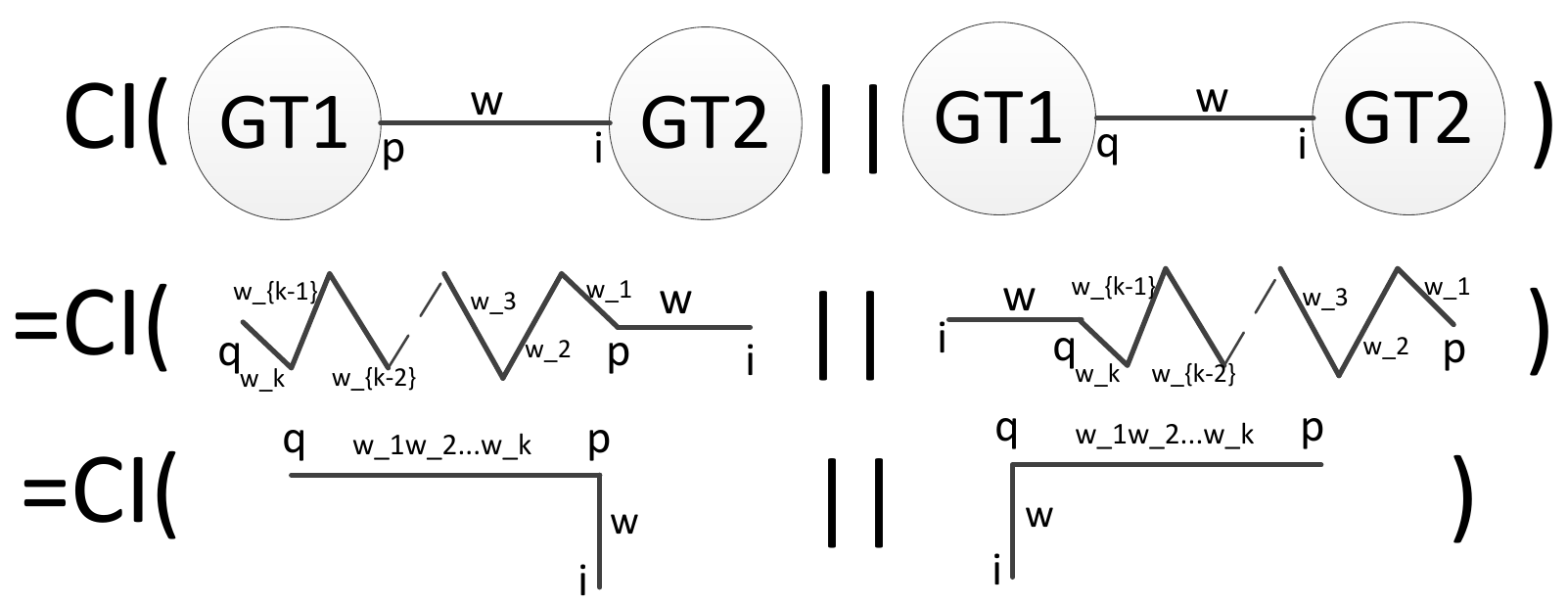}\\
    \caption{Examples of simplification}\label{6}
    \end{figure}

    This proposition is very important because we can use it to simplify the Chernoff information calculation of two Gaussian trees sharing the same local subtrees. If we use the proposition in opposite direction, we can find that cutting operation and merging operation can also keep  Chernoff information unchanged. So we can use these inverse operations repeatedly and reduce the complexity of Chernoff information calculation. Take Fig. \ref{6} as an example, the two big Gaussian trees have two big subparts which are totally the same, shown as subtree $G_{T_1}$ and $G_{T_2}$ represented by big circles. We can use cutting operation to delete the identical leaves of both trees and merging operation to delete the identical $2$-degree nodes in the path $p-q$. Then the Chernoff information calculation of big trees can be simplified into that of two small trees. The complexity is significantly reduced.

\subsection{Algebraic sufficient condition for $\lambda^*=0.5$}\label{0.5}

    It has been shown that $\lambda^*=0.5$ is a special equilibrium point for the calculation of Chernoff information, which means that the distinguishing cut-off point is in the "middle" of two trees. In this situation, $CI(\Sigma_1||\Sigma_2)=
    \frac{1}{2} \sum_i \ln\{ \sqrt{\lambda_i}+\frac{1}{\sqrt{\lambda_i}}\} -\frac{N}{2}\ln 2$, where $\{\lambda_i\}$ are the generalized eigenvalues of $\Sigma_1,\Sigma_2$.

     A sufficient but not necessary condition for $\lambda^*=0.5$ is stated below.

\begin{prop}\label{prop5}
    For two N-dim Gaussian joint distributions, $\mathbf{x}_1\sim N(0,\Sigma_1)$ and $\mathbf{x}_2\sim N(0,\Sigma_2)$, if there exists an involutory matrix $Q$ which meets condition $\Sigma_2=Q \Sigma_1 Q^T$, we can conclude $\lambda^*=0.5$.
\end{prop}

    Detail of the proof can be found in \ref{A3}.

    An involutory matrix is a matrix that is its own inverse, namely, $Q^2=I_N$. So the condition in Proposition \ref{prop5} is equivalent to
    $\{Q|Q^2=I\}\bigcap\{Q|\Sigma_2=Q \Sigma_1 Q^T\}\neq$ empty set.
    $Q$ means the linear transform of two trees' variables, namely, $\mathbf{x}_2=Q \mathbf{x}_1$. And $Q^2=I$ means $\mathbf{x}_2=Q \mathbf{x}_1,\mathbf{x}_1=Q \mathbf{x}_2$. The two trees $G_1,G_2$ are symmetrical on these linear transforms, which put the trees into a symmetric equivalence class.

     $\Sigma_2=Q \Sigma_1 Q^T\Longleftrightarrow\Sigma_2^{-1}=Q^T \Sigma_1^{-1} Q$. The topological structure of trees can be seen in $\Sigma_2^{-1}$ and $\Sigma_1^{-1}$. So $Q$ has to satisfy the structure constraint too.

\begin{figure}
      \centering
      \includegraphics[width=5cm]{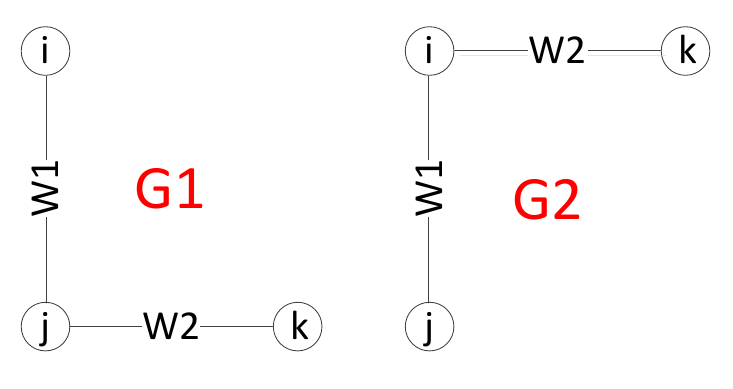}\\
      \caption{Special $3$-node case}\label{3}
    \end{figure}

    Here we consider a special case: $Q=Q^T=Q^{-1}$, where $Q$ is permutation matrix. We can define this operation as node-exchange operation. For the $3$-node case in Fig. \ref{3}, we can treat the operation  the exchange of node $i$ and $j$, namely $Q=\begin{pmatrix}0&1& 0\\1&0&0\\0&0&1\end{pmatrix}$. If two Gaussian trees have the same topology with different order of nodes, $\lambda^*$ between them equals to $0.5$.

    It is quite   special when $Q$ is a permutation matrix. So an immediate question from Proposition \ref{prop5} is how to find such a matrix $Q$ for arbitrarily case. Then we get  a proposition as below

\begin{prop}\label{prop6}
   For two covariance matrices $\Sigma_1$ and $\Sigma_2$, $Q$ meets condition $\Sigma_2=Q \Sigma_1 Q^T$ if and only if $Q=L_2 F L_1^{-1}$, where $L_1$ and $L_2$ are the Cholesky decomposition of $\Sigma_1$ and $\Sigma_2$, namely $\Sigma_1=L_1L_1^T, \Sigma_2=L_2L_2^T$, and $F$ is an arbitrary orthogonal matrix, namely $F^{-1}=F^T$.
\end{prop}

Detail of the proof can be found in   \ref{A4}. If we can find an involutory matrix in the set of $L_2 F L_1^{-1}$,  $\lambda^*=0.5$.

\section{Bottleneck: grafting operation}\label{section4}

    Former study tells us that overall Chernoff information in $M$-ary hypothesis testing is bottle-necked by the minimum pair-wise Chernoff information among the trees. So we want to learn the bottleneck of distinguishing Gaussian trees. Two Gaussian trees sharing the same local subtrees can be simplified into two smaller trees as shown in the former section, so we want to learn what kind of two trees can be simplified into the simplest trees.

    The simplest tree-pairs with the same entropy are two $3$-node trees. To keep the same entropy and focus on the topological changes, we make their edge weights unchanged, as shown in Fig. \ref{3}.
    The operation that can make two Gaussian trees simplified into this kind of $3$-node trees is  grafting operation. During grafting operation, we cut a subtree from another tree and paste it to another location, as shown in Fig. \ref{grafting}.
    Here $w_2$ is the weight of cut edge,  and $w_1$ is the weight of the path connecting the neighboring vertices of the moved node before and after the grafting operation.

\begin{figure}
        \centering
        \includegraphics[width=8.5cm]{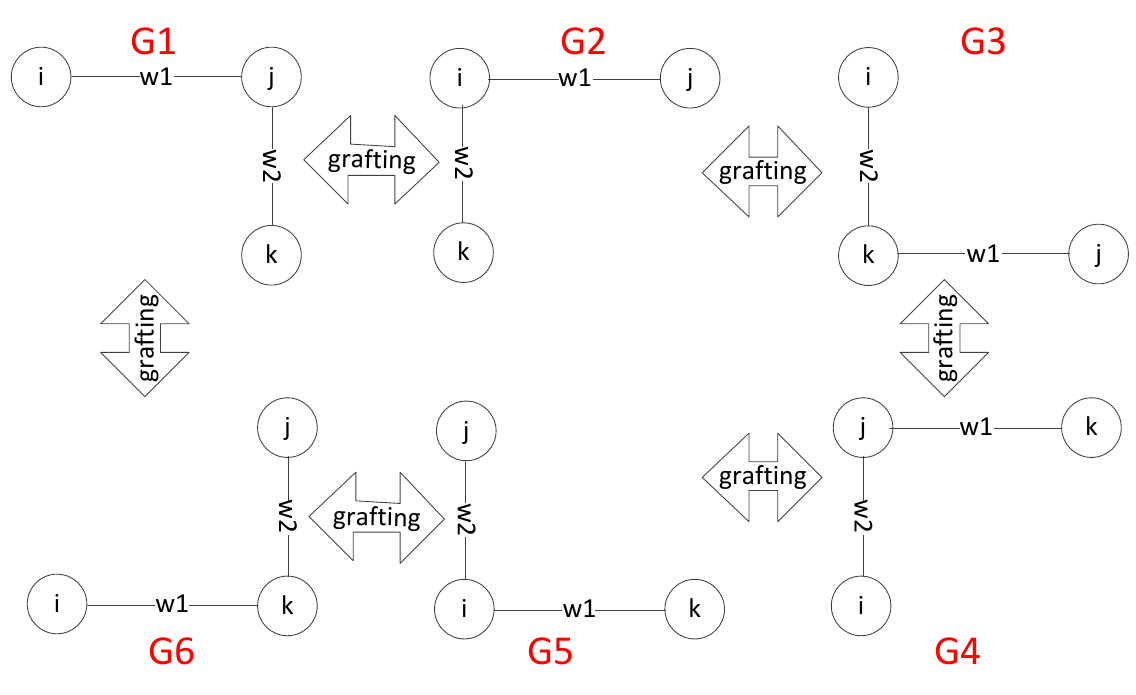}\\
        \caption{Different topology of $3$-node trees}\label{4}
    \end{figure}

    $3$-node trees with nodes $i,j,k$ and edge weights $w_1,w_2$ have six different patterns. And they can be connected by no more than $3$ grafting operations, as show in Fig. \ref{4}. So we treat grafting operation as the basic operation for bottleneck pair-wise Gaussian trees, when focusing on topological changes.

    Chernoff information between two big trees connected by grafting operation is the same with that of two $3$-node trees as shown in Fig. \ref{3}. So we study this special case first.

    Their covariance matrices are
    \begin{align}
    \Sigma_1=
    \begin{bmatrix} 1&w_1&w_1w_2\\w_1&1&w_2\\w_1w_2&w_2&1  \end{bmatrix}~~
    \Sigma_2=
    \begin{bmatrix} 1&w_1&w_2\\w_1&1&w_1w_2\\w_2&w_1w_2&1  \end{bmatrix}
    \end{align}

    \begin{prop}\label{16}
    For the two trees $\mathbf{G}_1$ and $\mathbf{G}_2$ in Fig. \ref{3}, the generalized eigenvalue of their correlation matrix $\Sigma_1$ and $\Sigma_2$, i.e.
    the eigen-values of the resulting matrix $S= \Sigma_1 \Sigma_2^{-1}$ is $\{1, \lambda_{max}, 1/\lambda_{max} \}$ with $\lambda_{max}$ determined by
    \begin{gather}
    \lambda_{max} = \frac{\sqrt{\beta} + w_2}{\sqrt{\beta} - w_2}
    \end{gather}
    \noindent where $\beta = w_2^2 + 2 \frac{1- w_2^2}{1-w_1}$
    \end{prop}

    We have taken Fig. \ref{3} as an example in section \ref{0.5}, and $Q$ which meets $\Sigma_2=Q \Sigma_1 Q^T$ is permutation matrix in this case. So $\lambda^*=0.5$ here due to Proposition \ref{prop5} and
    Chernoff information of $\mathbf{G}_1$ and $\mathbf{G}_2$ in Fig.~\ref{3} is
    \begin{align}
    CI(\Sigma_1||\Sigma_2)=\ln(\sqrt{\lambda_{max}}+\frac{1}{\sqrt{\lambda_{max}}})-\ln 2
    \end{align}

    We can use parameters $w_1$ and $w_2$ to express this Chernoff information, namely
    \begin{align}
    CI(\Sigma_1||\Sigma_2)=f(w_1,w_2)=\frac{1}{2}\log\left(1+\frac{1}{2}\frac{w_2^2}{1-w_2^2}(1-w_1)\right)\label{7}
    \end{align}

\section{Two Gaussian trees connected by more grafting operations}\label{section5}

     We are also interested in how Chernoff information changes when the trees are connected by more grafting operations. The answer of this question can tell us how Chernoff information changes when small differences merging into big difference. We anticipate Chernoff information to grow when small differences are added up into big difference, which does not always hold, as shown next. In this section, we start from a simple case, namely, two grafting operations. We label the Gaussian trees connected by two successive grafting operations as $G_1\leftrightarrow G_2 \leftrightarrow G_3$.

     There is a special situation in this case, when the two grafting operation involves the same cut edge. In this situation, the grafting chain will become a circle where $G_1$ can be obtained from $G_3$ by one grafting operation, or more specially, $G_1$ and $G_3$ can be the identical trees. This special situation is trival and we leave it out.

\subsection{Proposed comparison method}

\begin{figure}
    \centering
    \includegraphics[width=12cm]{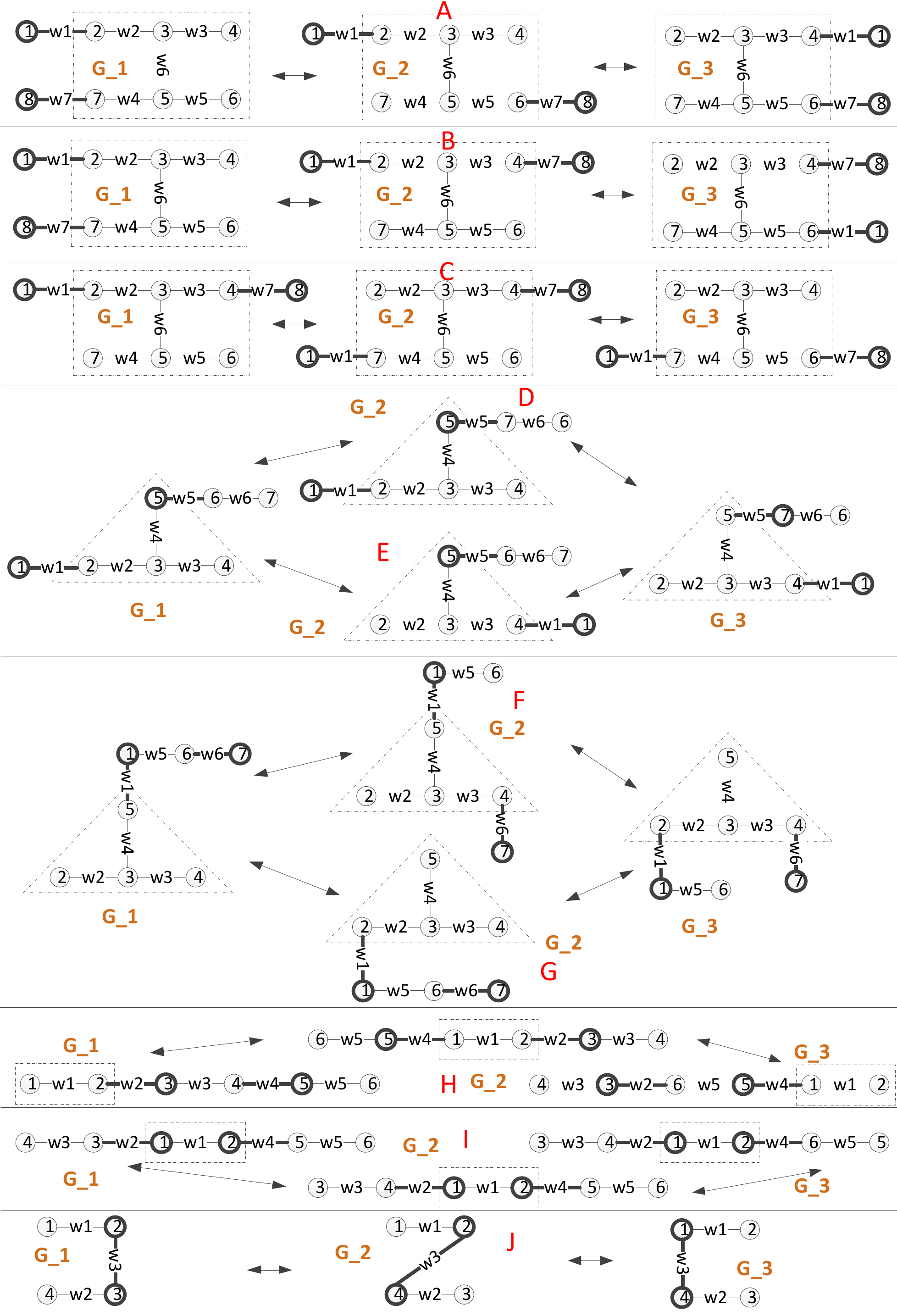}\\
    \caption{All possible cases for two successive grafting operations}\label{12}
    \end{figure}

    The two grafting operations can happen at different location, so the simplified trees of $G_1$ and $G_3$ have different patterns. Different simplified topology of these trees may have different characteristics. So
    we have to list all the possible patterns of $G_1\leftrightarrow G_2 \leftrightarrow G_3$ and analyze them respectively. The classification of such two successive operation is shown in Fig. \ref{12}.

\begin{figure*}
    \centering
    \includegraphics[width=12cm]{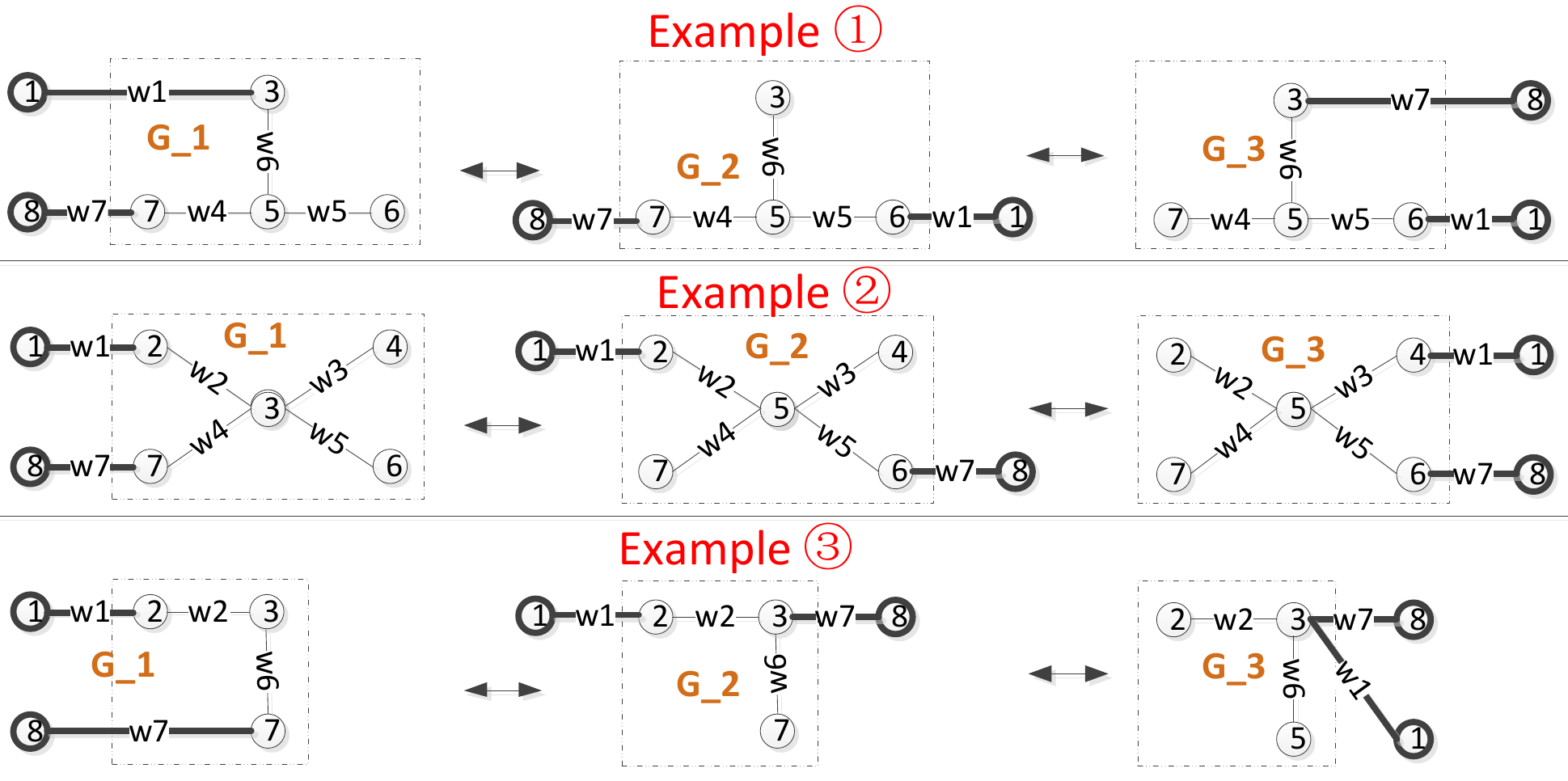}\\
    \caption{Example of Case A's special form }\label{9}
    \end{figure*}

    We give two definitions here, namely, anchor nodes and backbone of a grafting operation.
    Anchor nodes mean that we can simplify two big trees connected by grafting operation into two $3$-node trees with these nodes by using Proposition \ref{3}. For example, the anchor nodes of Fig. \ref{grafting} are $i,p,q$. Backbone of a grafting operation is the path from the original node to the target node, path $p-q$ in Fig. \ref{grafting} as an example.

    When two grafting operations are totally separate, there are $6$ anchor nodes, $3$ for each operation. Two of them are the leaves which will be removed. And the other four nodes are access points to backbones of the operations. The relative position of the four nodes remain the same during the operations. That is to say, the backbone in dashed box of Fig. \ref{12} is unchanged in two grafting operations.
    So we can make the classification with different size of backbones. We will start from backbone with $4$ access points.

    If there are four access points, the backbone can be simplified into a subtree with $6$ nodes, which has been shown in the dashed box of CASE A. The I-shaped subtree can express all the simplified relationship of $4$ access points. Then we go through all the possible operations from two access points to the other two access points and get CASE A,B,C. Note that node $6$ and node $7$ have the same relative position for node $2$ in the figure.

    If there are three access points, the backbone can be simplified into a subtree with $4$ nodes, which has been shown in the dashed box of CASE D. The star-shaped subtree can express all the possible relationship of $3$ access points. Here the access points $2,4,5$ have the same position in the star-shaped backbone.
    Then we go through all the possible two operations with access points $2,4,5$. Because the backbone lacks one access point, so one of the operation only has one access point, namely CASE D,E, or two operations share a same access point, namely CASE F,G.

    If there are two access points, the backbones can be simplified into a subtree with $2$ nodes, namely linearly $1-2$ subtree. If the operations are both from one access point to the other, we get CASE H. If all the operations are connected to one of the corresponding access points, we get CASE I.

    At last, there is a special case without a backbone, namely CASE J.

    In summary, we get ten cases of two successive grafting operations as shown in Fig. \ref{12}. The classification is a complete one, which is proved in  \ref{A6}.

    It is worth noting that we may make some edge weights in backbone approach $1$. That is to say, we can treat some connected nodes in the dashed frame to be the same node and make the simplified trees smaller. So the chains in Fig. \ref{9} can be treated as special forms of CASE A. Another important thing is that we can treat chain \textcircled{1} as a special case of CASE A rather than a new case with $3$ access points, although it actually has $3$ access points.

\subsection{Dependent and independent grafting operations}

    For the ten cases in Fig. \ref{12},  we can see that CASE F,G,H have something in common. The two successive grafting operations are overlapping with each other. Taking CASE F as an example, the backbone of the first operation is path $6-1-5-3-4$, which is destroyed in the second operation. Another example is CASE G. The backbone of the second grafting operation is $6-1-2-3-4$, which is just built in the first operation. So this kind of grafting operations are dependent on each other. If all the operations are not dependent, we call them independent grafting operations.

    \begin{figure}
    \centering
    \includegraphics[width=8cm]{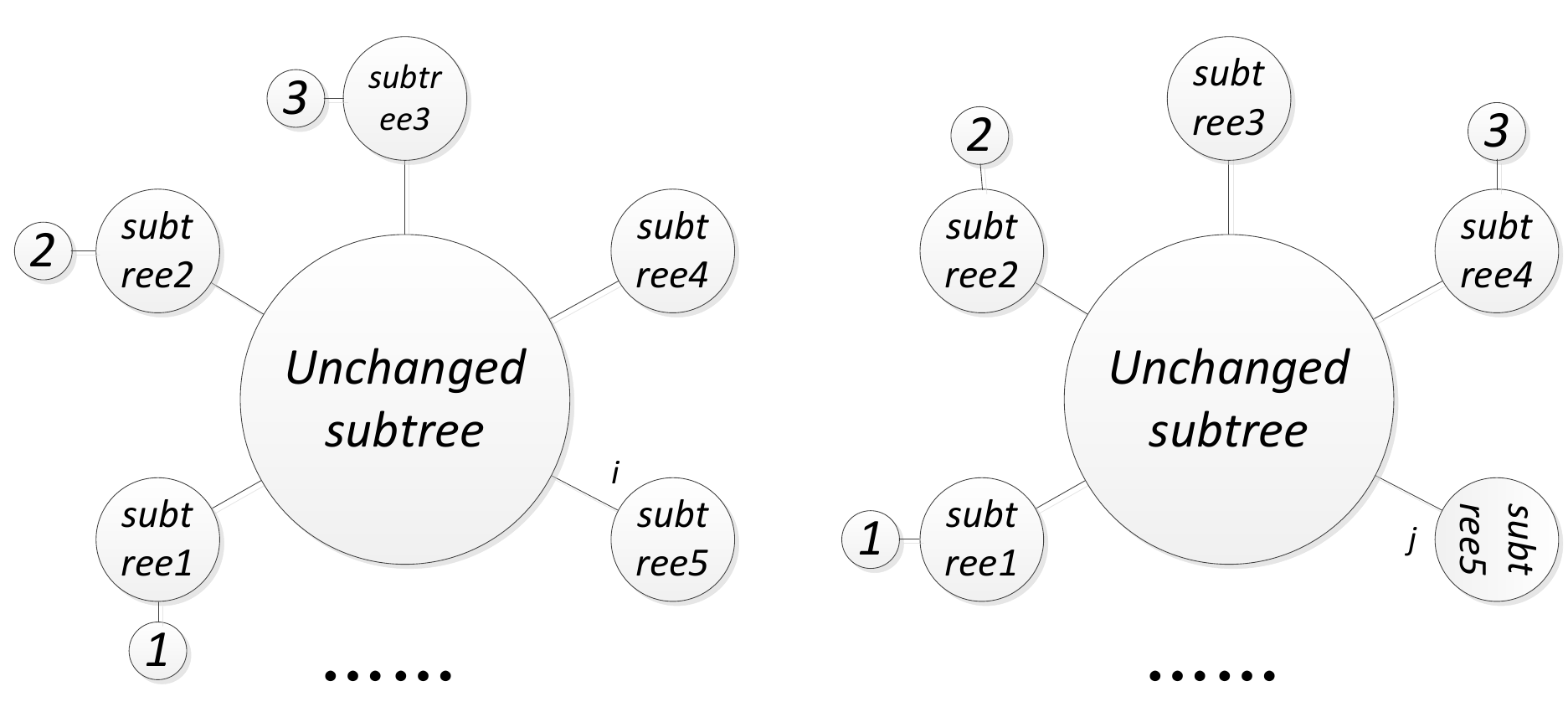}\\
    \caption{Independent grafting operations}\label{z}
    \end{figure}

    \begin{defn}
    If all the grafting operations can be divided into different subtrees, as shown in Fig. \ref{z}, then these grafting operations are independent. After regrouping all the nodes, the whole tree has star-shaped topology. The subtree in the center is unchanged during grafting operations. And grafting operations are involved in disjoint super leaf nodes of the star.
    \end{defn}

    In Fig. \ref{z}, we show $4$ independent grafting operations around the unchanged subtree. There are three types of grafting operations in the star-shaped topology. From left to right, the $1$-st, $2$-nd grafting operations belong to the first type, the $4$-th one belongs to the second type and the $3$-rd operation belongs to the third type. For the first type, we can cut a subtree, represented by a small circle with a number in it, from the super leaf node and paste it to another part of this super leaf node. In the second type, we can cut the unchanged subtree outside the super leaf node and paste it to another location in this super leaf node. But for the third type, we will cut a subtree, represented by a small circle with a number in it,   from a super leaf node and paste it to another super leaf node.  The third kind of grafting operation involves two super leaf nodes of the star while the first and second kinds of operation only involve one.

\subsection{Geometric sufficient conditions for $\lambda^*=0.5$}\label{chain}

    We have an  algebraic sufficient condition for $\lambda^*=1/2$ before. This sufficient condition gives us some insights on the relationship between generalized eigenvalues and covariance matrices. In this subsection, we  provide geometric sufficient conditions for judging whether $\lambda^*=1/2$.

    \begin{prop}\label{prop7}
    For two Gaussian trees connected by several independent grafting operations, $\lambda^*=1/2$ holds.
    \end{prop}

    CASE A,B,C,D,E,I in Fig. \ref{12} satisfy this condition, so $\lambda^*=1/2$ in these cases.

    The trees have the same number of nodes and the same entropy due to grafting operations. So $\lambda^*$ satisfies $tr(\Sigma_{\lambda^*}(\Sigma_1^{-1}-\Sigma_2^{-1}))=0$, which can be transformed from the definition formulas of $\lambda^*$. $tr(\Sigma_{\lambda^*}(\Sigma_1^{-1}-\Sigma_2^{-1}))$ is a summation formula with $4n$ term, where each $4$ terms are related to one single grafting operation. So we can deal with the terms respectively and prove $tr(\Sigma_{\frac{1}{2}}(\Sigma_1^{-1}-\Sigma_2^{-1}))=0$ eventually.
    More details can be found in   \ref{A7}.

    The successive operations can be treated as an operation chain, namely $T_1\leftrightarrow T_2 \leftrightarrow T_3 \leftrightarrow \dots \leftrightarrow T_{n+1}$. With fixed original tree and final tree, we have $n$ grafting operations. We can get different Gaussian tree chains with the same original tree and final tree by changing the order $\pi$ of the operations. So we have $n!$ different chains involving $2^n$ different trees, where all the super leaf nodes have two states: initial state and state after operation. For all the trees in the chains, $\lambda^*$ between two arbitrary trees, not just adjacent trees, is $0.5$ if these $n$ grafting operations are independent.

    The results in section \ref{0.5} show that $\lambda^*=1/2$ holds for adjacent trees in grafting chains because they can be simplified into two $3$-node trees where $Q$ is a permutation matrix. But $Q$ is not permutation matrix in most situations, for instance, $Q$ between $G_1$ and $G_3$ in all the cases of Fig. \ref{12} except CASE I and J. In these cases, $G_3$ can't be obtained from $G_1$ from node-exchange operation. So we can't judge if $\lambda^*=1/2$ holds by the results in section \ref{0.5}. So Proposition \ref{prop7} is much more powerful than that in most cases.

\section{Minimum Chernoff information in grafting chain}\label{section6}

    We have known that the overall error exponent in distinguishing a set of Gaussian trees is determined by the minimum pair-wise Chernoff information among the trees. So an immediate question is how to find out the minimum Chernoff information. We can find some inequality relationship or partial ordering to reduce the number of candidates.

    We first consider a grafting chain $T_1\leftrightarrow T_2 \leftrightarrow T_3$. Intuition tells us that $CI(T_1||T_3)$ is likely larger than $CI(T_1||T_2)$ and $CI(T_2||T_3)$, because the difference between $T_1$ and $T_3$ is the accumulation of $T_1-T_2$'s difference and   $T_2-T_3$'s difference. If the assumption is true, we only need to consider the adjacent pairs of Gaussian trees in the chain when looking for the minimum Chernoff information. Later results will tell us that it depends on the structure of the chain.

     \begin{prop}\label{combine-graph}
    Assuming two $N$-node Gaussian graphical models $\mathbf{G}_1=(V, E_1, W_1)$ and $\mathbf{G}_2=(V, E_2, W_2)$, we can draw two new Gaussian graphical models $\mathbf{G}_1'=(V/\{i\}, E_1', W_1'),\mathbf{G}_2'=(V/\{i\}, E_2', W_2')$ whose joint distributions are the same with the  joint distribution of $V/\{i\}$ nodes in $\mathbf{G}_1,\mathbf{G}_2$.  So $CI(\mathbf{G}_1'||\mathbf{G}_2')\leq CI(\mathbf{G}_1||\mathbf{G}_2)$.
    \end{prop}

    We can prove this proposition by transferring it into distribution field and use the Holder inequality.
    Details of proof can be found in  \cite{binglin}.

    This proposition tells us that Chernoff information can not increase if we ignore some nodes in both trees, for example, cutting the same leaves or merging the same $2$-degree nodes. Note that cutting operation and merging operation which we have defined before are special cases because these operations demand the cut or merged nodes have the same local structure in both trees before the operation. Cutting operation and merging operation precisely satisfy the equation condition for Proposition \ref{combine-graph}, which leads to Proposition \ref{prop4}.

    \begin{prop}\label{<}
    For the grafting chain $T_1\leftrightarrow T_2\leftrightarrow T_3 \leftrightarrow \dots \leftrightarrow T_n$ where all the grating operations in the chain are independent, we can conclude that $CI(T_i||T_j)\leq CI(T_p||T_q)$ if $p\leq i\leq j\leq q$.
    \end{prop}

    Detail of the proof can be found in  \ref{A9}.

     All the cases in Fig. \ref{12} except CASE F,G,H meet the condition, so $CI(G_1||G_3)\geq CI(G_1||G_2)$ and $CI(G_1||G_3)\geq CI(G_2||G_3)$ hold. The edge operation in \cite{tan2011large} is similar to the operation from $G_1$ to $G_3$ in CASE J, which involves only one edge and can be treated as two successive grafting operations. It is optimal in KL metric. But our study shows that this operation is not optimal when the problem uses Chernoff information metric. We can find a single grafting operation from $G_1$ to $G_2$ which can lead to a smaller Chernoff information than the edge operation from $G_1$ to $G_3$.

    If we want to find out the minimum Chernoff information among this set of independent grafting operations, we only need to try $n-1$ pairs of $T_i-T_{i+1}$, rather than all the  $\binom n 2$ pairs. The number of candidates is significantly reduced.

\subsection{Partial ordering}

    Former result provides us a partial ordering because we can't compare $CI(T_1||T_2)$ and $CI(T_2||T_3)$ in a simplest chain $T_1-T_2-T_3$ shown in Fig. \ref{12}. Taking CASE A as an example, we can turn $CI(T_1||T_2)$ and $CI(T_2||T_3)$ of this case into that of two different $3$-node cases. So $CI(T_1||T_2)=f(w_4w_5,w_7)$ and $CI(T_2||T_3)=f(w_2w_3,w_1)$, where $f(\cdot,\cdot)$ is defined in (\ref{7}). We can't compare them because they even don't share common parameters.

    We can only compare Chernoff information pairs $CI(T_i||T_j), CI(T_p||T_q)$ when $p\leq i\leq j\leq q$, and thus this result is a partial inequality, rather than a full ordering inequality.

\subsection{Cases without the partial ordering}

    At the same time, we also wonder whether the result suits for all the possible grafting chains $T_1\leftrightarrow T_2\leftrightarrow T_3 \leftrightarrow \dots \leftrightarrow T_n$. Taking CASE F in Fig. \ref{12} as an example, some special cases is shown in Table \ref{form2}. In this table, we can find that $CI(T_1||T_3)<CI(T_1||T_2)$ can hold in some special situations of CASE F.
    CASE G,H also have this kind of special situations. In CASE G, $CI(T_1||T_3)<CI(T_2||T_3)$ can hold. And in CASE H, either $CI(T_1||T_3)<CI(T_2||T_3)$ or $CI(T_1||T_3)<CI(T_1||T_2)$ can hold. But note that $CI(T_1||T_3)$ must not be the smallest one in all of these cases according to our numerical results.

\begin{table}
  \centering
  \begin{tabular}{|c|c|c|c|c|c|}
    \hline
    cases&$\lambda^*_{T_1||T_3}$&$\lambda$&
    $CI_{T_1||T_3}$&$CI_{T_1||T_2}$&$CI_{T_2||T_3}$\\
    \hline
     1&0.5191&$\begin{matrix}19.5746,
   0.0433,
   1.5439\\
   0.7642,
   1,
   1,
   1\end{matrix}$&0.8983&0.9142  &  0.0251\\
     \hline
     2&0.5073&$\begin{matrix}9.2341,
    0.1019,
    1.2982\\
    0.8185,
    1,
    1,
    1\end{matrix}$&0.5402&0.5418  &  0.0113 \\ \hline
    3&0.5254&$\begin{matrix}9.4328,
    1.653,
    0.0844\\
    0.7603,
    1,
    1,
    1\end{matrix}$&0.5982&0.6103  &  0.0392 \\
    \hline
     4&0.5082
&$\begin{matrix}5.0195,
    0.1863,
    1.2201\\
    0.8766,
    1,
    1,
    1\end{matrix}$&0.3102&0.3132&0.0056   \\
    \hline
    \end{tabular}
  \caption{Numerical cases dissatisfying Proposition \ref{<} in CASE F of Fig. \ref{12}}\label{form2}
\end{table}

    These special cases have two overlapping operations. So Proposition \ref{<} doesn't suit these cases. This is  a counter-intuitive result because more topological differences can't lead to larger Chernoff information between Gaussian trees.

    So we can only provide a partial ordering rather than a full ordering in this section. The problem of ordering is much more complex than what we have expected. For a grafting chain $T_1\leftrightarrow T_2 \leftrightarrow T_3$, the Chernoff information between $T_1$ and $T_3$ may be larger than that between $T_1$ and $T_2$, even though the difference between $T_1$ and $T_2$ seems smaller. Here we only consider topological difference, rather than parameterized difference, between Gaussian trees.
    So topological difference is not the only contribution factor affecting such comparisons\cite{tan2011large,jog2015model}.

\section{Dimension reduction}\label{section7}

    The situations in the former section are all about full observation cases, where we can observe all Gaussian variables in the trees each time.
    But in practice, we may have some constraint on observation costs, which prompts us to reduce the dimension of observation vectors  in order to meet such  constraints.

    The simplest mapping from $N$-dim vector $X$ to low $N_O$-dim output vector $Y$ is linear transformation (LT), namely, $Y=AX$, where $A$ is a $N_O\times N$ mapping matrix and $1\leq N_O\leq N$. The question is how to find the optimal mapping to maximize  Chernoff information of output distributions.

    The more dimension of data we have access to, the larger optimal Chernoff information we can get. Due to the LT operation, we have to drop some dimension of data, which will lead to smaller optimal Chernoff information.

    \begin{prop}\label{>1}
    Assume two $N$-dim distributions $\mathbf{X}_{1}$ and $\mathbf{X}_{2}$ on two graphs. We can use two different observation matrices to get different outputs $\mathbf{Y}_p=\mathbf{P}_{p\times N}\mathbf{X}$ and $\mathbf{Y}_q=\mathbf{Q}_{q\times N}\mathbf{X}$ where $q<p\leq N$ and $\mathbf{P}^*,\mathbf{Q}^*$ are the optimal matrix  under the observation dimension constraint. So $CI(\mathbf{Y}_p^{(1)*}||\mathbf{Y}_p^{(2)*})\geq CI(\mathbf{Y}_q^{(1)*}||\mathbf{Y}_q^{(2)*})$.
    \end{prop}

    Detail of proof can be found in \cite{binglin}.

For two $N$-node $0$-mean Gaussian graphs $G_1$ and $G_2$ on random variables $\mathbf{x}$, whose covariance matrices are $\mathbf{\Sigma}_1$ and $\mathbf{\Sigma}_2$, we can use an inverse linear transformation matrix $\mathbf{P}$ to transform them to $\mathbf{x}'=\mathbf{P}\mathbf{x}$ whose covariance matrices $\mathbf{\Sigma}_1'$ and $\mathbf{\Sigma}_2'$ are diagonal and related to the generalized eigenvalues of $\mathbf{\Sigma}_1$ and $\mathbf{\Sigma}_2$.

$\mathbf{\Sigma}_1$ and $\mathbf{\Sigma}_2$ are real symmetric positive definite matrices, so the eigenvalues of $\mathbf{\Sigma}_1\mathbf{\Sigma}_2^{-1}$ are all positive, as shown in  \ref{AA1}.
The eigenvalue decomposition of $\mathbf{\Sigma}_1\mathbf{\Sigma}_2^{-1}$ is $\mathbf{Q}\mathbf{\Lambda} \mathbf{Q}^{-1}$, where $\mathbf{Q}$ is an $N\times N$ matrix and $\mathbf{\Lambda}=Diag(\{\lambda_i\})$ is a diagonal matrix of eigenvalues, in which we put multiple eigenvalues adjacent. $\{\lambda_i\}$ are the eigenvalues of $\mathbf{\Sigma}_1\mathbf{\Sigma}_2^{-1}$, namely, the generalized eigenvalues of $\mathbf{\Sigma}_1$ and $\mathbf{\Sigma}_2$.
Note that $\mathbf{Q}$ may be non-orthogonal when $\mathbf{\Sigma}_1\mathbf{\Sigma}_2^{-1}$ isn't symmetric.

\begin{prop}\label{prop0}
For two $N$-node $0$-mean Gaussian graphs $G_1$ and $G_2$ whose covariance matrices are $\mathbf{\Sigma}_1$ and $\mathbf{\Sigma}_2$ respectively,
we can construct a linear transformation matrix $\mathbf{P}={\left(\mathbf{Q}^{-1}\mathbf{\Sigma}_2{(\mathbf{Q}^{-1})}^T\right)}^{-\frac{1}{2}}\mathbf{Q}^{-1}$ and thus
\begin{align}
\mathbf{\Sigma}_2'=&\mathbf{P}\mathbf{\Sigma}_2\mathbf{P}^T=\mathbf{I}_N\\
\mathbf{\Sigma}_1'=&\mathbf{P}\mathbf{\Sigma}_1\mathbf{P}^T=\mathbf{\Lambda}
\end{align}
where eigenvalue decomposition of $\mathbf{\Sigma}_1\mathbf{\Sigma}_2^{-1}$ is $\mathbf{Q}\mathbf{\Lambda} \mathbf{Q}^{-1}$.
\end{prop}

The proof of Proposition \ref{prop0} is shown in \ref{AA1}.

 We can treat $\mathbf{\Sigma}_1'$ and $\mathbf{\Sigma}_2'$ as two  graphs $G_1'$ and $G_2'$ on $\mathbf{x}'$ obtained from $G_1$ and $G_2$ by inverse linear transformation $\mathbf{P}$.
$G_1'$ and $G_2'$ are graphs with $N$ independent variables.

The distances of $G_1'$ and $G_2'$ are as follows. The distances between $G_1$ and $G_2$ are the same with the distances of $G_1'$ and $G_2'$ just because $\mathbf{P}$ is inverse.

\begin{align}
D(\mathbf{\Sigma}_1||\mathbf{\Sigma}_2)=&D(\mathbf{\Sigma}_1'||\mathbf{\Sigma}_2')=\frac{1}{2}\sum_i \left( -\ln\lambda_i+\lambda_i-1\right)\\
D(\mathbf{\Sigma}_2||\mathbf{\Sigma}_1)=&D(\mathbf{\Sigma}_2'||\mathbf{\Sigma}_1')=\frac{1}{2}\sum_i \left( \ln\lambda_i+\frac{1}{\lambda_i}-1\right)\\
D(\mathbf{\Sigma}_\lambda||\mathbf{\Sigma}_1)=&D(\mathbf{\Sigma}_\lambda'||\mathbf{\Sigma}_1')\nonumber\\=\frac{1}{2}\sum_i&\left(\ln\left(\lambda+(1-\lambda)\lambda_i\right)+\frac{1}{\lambda+(1-\lambda)\lambda_i}-1\right)\\
D(\mathbf{\Sigma}_\lambda||\mathbf{\Sigma}_2)=&D(\mathbf{\Sigma}_\lambda'||\mathbf{\Sigma}_2')\nonumber\\=\frac{1}{2}\sum_i&\left(\ln\frac{\lambda+(1-\lambda)\lambda_i}{\lambda_i}+\frac{\lambda_i}{\lambda+(1-\lambda)\lambda_i}-1\right)\\
CI(\mathbf{\Sigma}_1||\mathbf{\Sigma}_2)=&CI(\mathbf{\Sigma}_1'||\mathbf{\Sigma}_2')=D(\mathbf{\Sigma}_{\lambda^*}'||\mathbf{\Sigma}_1')=D(\mathbf{\Sigma}_{\lambda^*}'||\mathbf{\Sigma}_2')\label{CI}
\end{align}
where $\mathbf{\Sigma}_\lambda^{-1}=\mathbf{\Sigma}_1^{-1}\lambda+\mathbf{\Sigma}_2^{-1}(1-\lambda)$ and ${\mathbf{\Sigma}_\lambda'}^{-1}={\mathbf{\Sigma}_1'}^{-1}\lambda+{\mathbf{\Sigma}_2'}^{-1}(1-\lambda)$. $\mathbf{\Sigma}_\lambda'$ is also a diagonal matrix.
$\lambda^*$ in (\ref{CI}) is the unique root of $D(\mathbf{\Sigma}_{\mathbf{\lambda}^*}'||\mathbf{\Sigma}_1')=D(\Sigma_{\mathbf{\lambda}^*}'||\mathbf{\Sigma}_2')$, namely,
$\sum_i\left(\frac{1-\lambda_i}{\lambda^*+(1-\lambda^*)\lambda_i}+\ln\lambda_i\right)=0$.

In these new space, $x_i'$, the $i$-th variable of $\mathbf{x}'$, follows $N(0,1)$ in hypothesis $2$ and $N(0,\lambda_i)$ in hypothesis $1$. If $\lambda_i$ is farther from $1$, this dimension can provide more information for classification than other dimensions.

Assume that $m$ of all the $N$ eigenvalues $\{\lambda_i\}$ are greater than $1$ and the other $N-m$ eigenvalues are no more than $1$.
If we want to reduce the observation dimension from $N$ to $N_O$, we choose the dimensions of $\mathbf{x}'$ corresponding to the first $k$ rank and last $N_O-k$ rank of $\{\lambda_i\}$, where $N_O+m-N\leq k\leq m$ and $k\geq0$.
We can choose a sub-matrix of $\mathbf{\Sigma}_1'$ and $\mathbf{\Sigma}_2'$ corresponding to the $N_O$ chosen eigenvalues as the result of dimension reduction. The $N_O\times N$ linear transformation matrix $\mathbf{A}_k$ is the corresponding $N_O$ rows of $\mathbf{P}$ corresponding to the chosen eigenvalues.

\begin{prop}\label{DR}
    $\mathbf{A}^*$ is the optimal $N_O\times N$ linear transformation matrices to maximize the Chernoff information in transformed space, namely
    $\mathbf{A}^*=\arg \max_{\mathbf{A}_{N_O\times N}} CI(\hat{\mathbf{\Sigma}}_1||\hat{\mathbf{\Sigma}}_2)$ where $\hat{\mathbf{\Sigma}}_i=\mathbf{A}^*\mathbf{\Sigma}_i{\mathbf{A}^*}^T$ for $i=1,2$. So $\mathbf{A}^*\in\{\mathbf{A}_k|N_O+m-N\leq k\leq m, k\geq0\}$.
    \end{prop}

Proof of proposition \ref{DR} can be seen in \ref{AA7} and this proposition ensure the optimality of our method.

The observation is $\mathbf{y}=\mathbf{A}^*\mathbf{x}$ and the covariance matrices of $\mathbf{y}$ in two hypothesis are
    \begin{align}
\mathbf{\Sigma}_2''=\mathbf{A}^*\mathbf{\Sigma}_2{\mathbf{A}^*}^T=\mathbf{I}_{N_O}\\
\mathbf{\Sigma}_1''=\mathbf{A}^*\mathbf{\Sigma}_1{\mathbf{A}^*}^T=Diag(\{\mu_i\})
\end{align}
where $\mathbf{\Sigma}_1''$ and $\mathbf{\Sigma}_2''$ are $N_O\times N_O$ diagonal matrices and $\{\mu_1,\mu_2,\dots,\mu_{N_O}\}$ (including multiple eigenvalues) are $N_O$ chosen eigenvalues.

\begin{align}
&D(\mathbf{\Sigma}_\lambda''||\mathbf{\Sigma}_1'')\nonumber\\
=&\frac{1}{2}\sum_i\left(\ln\left(\lambda+(1-\lambda)\mu_i\right)+\frac{1}{\lambda+(1-\lambda)\mu_i}-1\right)\\
&D(\mathbf{\Sigma}_\lambda''||\mathbf{\Sigma}_2'')\nonumber\\
=&\frac{1}{2}\sum_i\left(\ln\frac{\lambda+(1-\lambda)\mu_i}{\lambda_i}+\frac{\lambda_i}{\lambda+(1-\lambda)\mu_i}-1\right)\\
&CI(\mathbf{\Sigma}_1''||\mathbf{\Sigma}_2'')=\max_{\lambda\in[0,1]}\min\{D(\mathbf{\Sigma}_\lambda''||\mathbf{\Sigma}_1''),D(\mathbf{\Sigma}_\lambda''||\mathbf{\Sigma}_2'')\}\label{CI2}
\end{align}
Equation (\ref{CI2}) is another definition of Chernoff information\cite{cover2012elements}. The distances are only related to the eigenvalues we have chosen.

\section{Conclusion}\label{section8}
In this paper,
we have shown the relationship between Chernoff information and generalized eigenvalues and
how local changes in topology affect the capacity of distinguishing two resulting Gaussian
trees measured by Chernoff information. Moreover, we have shown how Chernoff information changes when small topological differences merging into bigger ones. We also study the dimension reduction case and show that optimal Chernoff information  is also determined by generalized eigenvalues.

\section*{Acknowledgments}
This work is  supported in part by the National Science
Foundation (USA) under Grant No. 1320351, and the National Natural Science Foundation of China under Grant 61673237.
\section*{References}
\bibliographystyle{elsarticle-num}
\bibliography{Bibliography}

\appendix

\section{Proof of proposition~\ref{prop1}}\label{A1}

For two N-dim Gaussian joint distribution, $\mathbf{x}_1\sim N(0,\Sigma_1)$ and $\mathbf{x}_2\sim N(0,\Sigma_2)$. $CI(\Sigma_1||\Sigma_2)=\frac{1}{2}\log\frac{|\Sigma_1|}{|\Sigma_{\lambda^*}|}=\frac{1}{2}\log\frac{|\Sigma_2|}{|\Sigma_{\lambda^*}|}$,
   when $|\Sigma_1|=|\Sigma_2|$ and
 $\lambda^*$ satisfying $tr(\Sigma_1^{-1}\Sigma_{\lambda^*})=tr(\Sigma_2^{-1}\Sigma_{\lambda^*})$.

    \begin{align}
    CI(\Sigma_1||\Sigma_2)=&\frac{1}{2}\ln\frac{|\Sigma_1|}{|\Sigma_{\lambda^*}|}
    \nonumber\\=& \frac{1}{2}\ln(|\Sigma_1||\lambda^*\Sigma_1^{-1}+(1-\lambda^*)\Sigma_2^{-1}|)
    \nonumber\\=&
    \frac{1}{2}\ln|\lambda^*I+(1-\lambda^*)\Sigma_1\Sigma_2^{-1}|
    \nonumber\\\overset{(a)}{=}& \frac{1}{2} \sum_i \left\{\ln\{(1-\lambda^*) \lambda_i+\lambda^*\}\right\}
    \nonumber\\\overset{(b)}{=}&
    \frac{1}{2} \sum_i \left\{\ln\{(1-\lambda^*) \sqrt{\lambda_i}+\frac{\lambda^*}{\sqrt{\lambda_i}}\}\right\}
    \end{align}

    $\{\lambda_i\}$ are the eigenvalues of $\Sigma_1\Sigma_2^{-1}$, namely, the generalized eigenvalues of $\Sigma_1$ and $\Sigma_2$.
    $(a)$ is true because $\{(1-\lambda^*) \lambda_i+\lambda^*\}$ are the eigenvalues of $\lambda^*I+(1-\lambda^*)\Sigma_1\Sigma_2^{-1}$.  $(b)$ is true because $\prod_{i}\lambda_i=|\Sigma_1\Sigma_2^{-1}|=1$.

\begin{align}
tr(\Sigma_2^{-1}\Sigma_{\lambda^*}) &=\sum_{i} {Eig}_i(\Sigma_2^{-1}\Sigma_{\lambda^*})
\nonumber\\&=\sum_{i} \frac{1}{{Eig}_i(\Sigma_{\lambda^*}^{-1}\Sigma_2)}
\nonumber\\&=\sum_{i} \frac{1}{{Eig}_i\left(\lambda^*\Sigma_{1}^{-1}\Sigma_2+(1-\lambda^*)I\right)}
\nonumber\\&=\sum_{i} \frac{1}{\lambda^*\frac{1}{\lambda_i}+(1-\lambda^*)}
\nonumber\\&=\sum_{i} \frac{\lambda_i}{\lambda^*+(1-\lambda^*)\lambda_i}\\
tr(\Sigma_1^{-1}\Sigma_{\lambda^*}) &=\sum_{i} Eig_i(\Sigma_1^{-1}\Sigma_{\lambda^*})
\nonumber\\&=\sum_{i} \frac{1}{Eig_i(\Sigma_{\lambda^*}^{-1}\Sigma_1)}
\nonumber\\&=\sum_{i} \frac{1}{Eig_i\left(\lambda^*I+(1-\lambda^*)\Sigma_{2}^{-1}\Sigma_1\right)}
\nonumber\\&=\sum_{i} \frac{1}{\lambda^*+(1-\lambda^*)\lambda_i}
\end{align}
where $Eig_i(A)$ is the $i$-th eigenvalue of matrix $A$. $\{\lambda_i\}$ are the eigenvalues of $\Sigma_1\Sigma_2^{-1}$, so $\{\lambda_i\}$ are the eigenvalues of $\Sigma_2^{-1}\Sigma_1$ and $\{1/\lambda_i\}$ are the eigenvalues of $\Sigma_1^{-1}\Sigma_2$.

So $\lambda^*$ satisfies
\begin{align}
\sum_{i} \frac{1}{\lambda^*+(1-\lambda^*)\lambda_i}=\sum_{i} \frac{\lambda_i}{\lambda^*+(1-\lambda^*)\lambda_i}
\end{align}

\section{Proof of proposition~\ref{prop2}}\label{A2}

    We use the equation of block matrix to prove this proposition. We deal with adding operation and division operation in different subsections.

\subsection{Proof on adding operation case}

    Adding operation is shown in Fig. {\ref{adding}}. Assume $G_1$ and $G_2$ have node $1,\dots,N$, and their covariance matrices are

    $G_1$: $\Sigma_1=\begin{bmatrix} D&\alpha\\ \alpha^T&1 \end{bmatrix}$

    $G_2$: $\Sigma_2$, $\Sigma_2^{-1}=\begin{bmatrix} A_2&B_2\\B_2^T&a_2 \end{bmatrix}$.

     Without loss of generality, the new edge is $e_{N,N+1}$ with parameter $w$. So new covariance matrices after adding operation are

     \begin{align}
     \Sigma_1'=\begin{bmatrix} D&\alpha&w\alpha\\ \alpha^T&1&w\\w\alpha^T&w&1 \end{bmatrix}\\
     \Sigma_2'^{-1}=
     \begin{bmatrix} A_2&B_2&0\\B_2^T&a_2+\frac{w^2}{1-w^2}&\frac{-w}{1-w^2}\\
     0&-\frac{w}{1-w^2}&\frac{1}{1-w^2} \end{bmatrix}
     \end{align}

     The generalized eigenvalues of $\Sigma_1$ and $\Sigma_2$ are the roots of $|\lambda I -\Sigma_1\Sigma_2^{-1}|=0$.

     \begin{align}
     \Sigma_1\Sigma_2^{-1}=&\begin{bmatrix} D&\alpha\\ \alpha^T&1 \end{bmatrix}\begin{bmatrix} A_2&B_2\\B_2^T&a_2 \end{bmatrix}\nonumber\\
     =&
     \begin{bmatrix} DA_2+\alpha B_2^T&DB_2+a_2\alpha\\ \alpha^TA_2+B_2^T& \alpha^TB_2+a_2 \end{bmatrix}\\
     \Sigma_1'\Sigma_2'^{-1}=&
     \begin{bmatrix}
     DA_2+\alpha B_2^T&DB_2+a_2\alpha&0\\
     \alpha^TA_2+B_2^T& \alpha^TB_2+a_2 &0\\
     \ast&\ast&1
     \end{bmatrix}\\
     |\lambda I -\Sigma_1'\Sigma_2'^{-1}|=&(\lambda-1)|\lambda I -\Sigma_1\Sigma_2^{-1}|
     \end{align}
     So new trees have an extra generalized eigenvalue $1$ without changing other eigenvalues.

\subsection{Proof on division operation case}

    Division operation is shown in Fig. {\ref{division}}. Assume that $G_1$ and $G_2$ have node $1,\dots,N$,  and the connecting points are $p=1,q=N$ without loss of generality.

    If we set edge $e_{1N}$ in $G_2$ to be $0$, we can treat it as a new $N$-node tree, whose covariance matrix is $A^{-1}$.
    We divide A as
    \begin{align}
    A=\begin{bmatrix} x&X^T&0\\X&Z&Y\\0&Y^T&y \end{bmatrix}
    \end{align}
    where $X,Y$ are $(N-2)\times1$ matrices and $Z$ is $(N-2)\times(N-2)$ matrix.

    In $G_1$ and $G_1'$, we define the column node $1$ to nodes $2:N-1$ to be $\alpha_1$(setting the respective rows to the nodes in $G_{T_{12}}$ $0$) and the column node $N$ to nodes $2:N-1$ to be $\alpha_2$(setting the respective rows to the nodes in $G_{T_{11}}$ $0$). Then in $G_1$ and $G_1'$, the column from node $1$ to nodes $2:N-1$ is $\alpha_1+w_1w_2\alpha_2$ because node $1$ can reach the nodes in $G_{T_{11}}$ directly, but reach the nodes in $G_{T_{12}}$ through node $N$ with path weighted by $w_{1N}=w_1w_2$. And the column from node $N$ to node $2:N-1$ is $w_1w_2\alpha_1+\alpha_2$ because node $N$ can reach the nodes in $G_{T_{12}}$ directly, but reach the nodes in $G_{T_{11}}$ through node $1$ with path weighted by $w_{1N}=w_1w_2$. In $G_1'$, the column from node $N+1$ to node $2:N-1$ is $w_1\alpha_1+w_2\alpha_2$ because node $N+1$ can reach the nodes in $G_{T_{11}}$ through node $1$ with path $e_{1,N+1}$ weighted by $w_{1}$, and reach the nodes in $G_{T_{12}}$ through node $N$ with path $e_{N,N+1}$ weighted by $w_{2}$.

    The covariance matrices of these trees are as shown below:

$G_1$: $\Sigma_{11}=
    \begin{bmatrix} 1&\alpha_1^T+w_1w_2\alpha_2^T&w_1w_2\\
    \alpha_1+w_1w_2\alpha_2 &P&w_1w_2\alpha_1+\alpha_2 \\
    w_1w_2&w_1w_2\alpha_1^T+\alpha_2^T &1\end{bmatrix}$

    $G_2$: $\Sigma_{12}$, $\Sigma_{12}^{-1}=\begin{bmatrix} x+\frac{w_1^2w_2^2}{1-w_1^2w_2^2}&X^T&-\frac{w_1w_2}{1-w_1^2w_2^2}\\X&Z&Y\\-\frac{w_1w_2}{1-w_1^2w_2^2}&Y^T&y+\frac{w_1^2w_2^2}{1-w_1^2w_2^2} \end{bmatrix}$.

    $G_1'$: $\Sigma_{21}=\begin{bmatrix} 1&\alpha_1^T+w_1w_2\alpha_2^T&w_1w_2&w_1\\
    \alpha_1+w_1w_2\alpha_2 &P&w_1w_2\alpha_1+\alpha_2 &w_1\alpha_1+w_2\alpha_2\\
    w_1w_2&w_1w_2\alpha_1^T+\alpha_2^T &1&w_2\\
    w_1&w_1\alpha_1^T+w_2\alpha_2^T&w_2&1 \end{bmatrix}$

    $G_2'$: $\Sigma_{22}$,

    $\Sigma_{22}^{-1}=\begin{bmatrix} x+\frac{w_1^2}{1-w_1^2}&X^T&0&-\frac{w_1}{1-w_1^2}\\
    X&Z&Y&0\\
    0&Y^T&y+\frac{w_2^2}{1-w_2^2}&-\frac{w_2}{1-w_2^2}\\
    -\frac{w_1}{1-w_1^2}&0&-\frac{w_2}{1-w_2^2}&\frac{1}{1-w_1^2}+\frac{w_2^2}{1-w_2^2} \end{bmatrix}$.

     The generalized eigenvalues of $\Sigma_1$ and $\Sigma_2$ are the roots of $|\lambda I -\Sigma_1\Sigma_2^{-1}|=0$.
     \begin{align}
     \Sigma_{11}\Sigma_{12}^{-1}=&\Sigma_{11}A+
     \begin{bmatrix} 0&0&-w_1w_2\\
     -w_1w_2\alpha_2&0&-w_1w_2\alpha_1\\
     -w_1w_2&0&0 \end{bmatrix}\\
     \Sigma_{21}\Sigma_{22}^{-1}=&
     \begin{bmatrix}
     \Sigma_{11}\Sigma_{12}^{-1}&\boldsymbol{0}\\
     \ast&1
     \end{bmatrix}\\
     |\lambda I -\Sigma_{21}\Sigma_{22}^{-1}|=&(\lambda-1)|\lambda I -\Sigma_{11}\Sigma_{12}^{-1}|
     \end{align}
     So new trees have an extra generalized eigenvalue $1$ without changing other eigenvalues.

\section{Proof of proposition~\ref{prop5}}\label{A3}

Assuming $\Sigma_2=Q\Sigma_1Q^T$, $Q^{-1}=Q$,
\begin{align}
    Q^T=&{(Q^T)}^{-1}\\
    \Sigma_1\Sigma_2^{-1}=&\Sigma_1{(Q^T)}^{-1}\Sigma_1^{-1}Q^{-1}\\
    \Sigma_2\Sigma_1^{-1}=&Q \Sigma_1 Q^T \Sigma_1^{-1}\\
    Q\Sigma_1\Sigma_2^{-1}Q^{-1}=&Q\Sigma_1{(Q^T)}^{-1}\Sigma_1^{-1}Q^{-1}Q^{-1}\nonumber\\=
    &Q\Sigma_1Q^T\Sigma_1^{-1}=\Sigma_2\Sigma_1^{-1}
    \end{align}

    So $Q\Sigma_1\Sigma_2^{-1}Q^{-1}=\Sigma_2\Sigma_1^{-1}$. $\Sigma_1\Sigma_2^{-1}$ and $\Sigma_2\Sigma_1^{-1}$ are both mutual similar matrices and mutually inverse matrices. Then the eigenvalues of $\Sigma_1\Sigma_2^{-1}$ appear in reciprocal pairs. That is to say, the generalized eigenvalues of $\Sigma_1$ and $\Sigma_2$ appear in reciprocal pairs. $1/a$ is the generalized eigenvalues of $\Sigma_1$ and $\Sigma_2$ if $1/a$ is the generalized eigenvalues. So
    \begin{align}
    \sum_{i} \frac{1}{0.5+(1-0.5)\lambda_i}&=\sum_{i} \frac{1}{0.5+(1-0.5)\frac{1}{\lambda_i}}\nonumber\\
    &=\sum_{i} \frac{\lambda_i}{0.5+(1-0.5)\lambda_i}
    \end{align}

     So we can conclude  $\lambda^*=0.5$ from Proposition \ref{prop1}.

\section{Proof of proposition~\ref{prop6}}\label{A4}

In order to prove Proposition \ref{prop6}, we have to prove Proposition \ref{t} as below:
    \begin{prop}\label{t}
    $P_2P_2^T=P_1P_1^T\Longleftrightarrow P_2=P_1F$ where $F$ is any orthogonal matrix.
    \end{prop}

We prove Proposition \ref{t} in two direction:

    1)$P_2=P_1F\longrightarrow  P_2P_2^T=P_1P_1^T$

     $P_2P_2^T=P_1FF^TP_1^T=P_1P_1^T$

    2)$P_2P_2^T=P_1P_1^T\longrightarrow P_2=P_1F$

    We define $P_2P_2^T=P_1P_1^T=A=U\Lambda U^T$ where $A$ is a real symmetric matrix and $U\Lambda U^T$  is its eigenvalue decomposition. $D=\Lambda^{\frac{1}{2}}=diag\{\sqrt{\lambda_i}\}$.
    So $P_1=UDV_1,P_2=UDV_2$ where $V_1=D^{-1}U^TP_1,V_2=D^{-1}U^TP_2$ are orthogonal matrices. So $P_2=P_1V_1^TV_2=P_1F$.

    So Proposition \ref{t} is proved.

   We use Cholesky decomposition to deal with the  covariant matrices $\Sigma_1$ and $\Sigma_2$.

   \begin{align}
   \Sigma_1=&L_1L_1^T\\
   \Sigma_2=L_2L_2^T=&Q \Sigma_1 Q^T=(QL_1){(QL_1)}^T\label{5}
   \end{align}
   where $L_1,L_2$ are specific lower triangular matrix corresponding to $\Sigma_1$ and $\Sigma_2$.
   \begin{align}
   L_2{L_2}^T=&(QL_1){(QL_1)}^T\\
   L_2F=&QL_1\\
   Q=&L_2FL_1^{-1}
   \end{align}
   where $F$ is an arbitrary orthogonal matrix.

   So Proposition \ref{prop6} holds.

\section{Classification for two successive grafting operations}\label{A6}

\begin{figure}
    \centering
    \includegraphics[width=8cm]{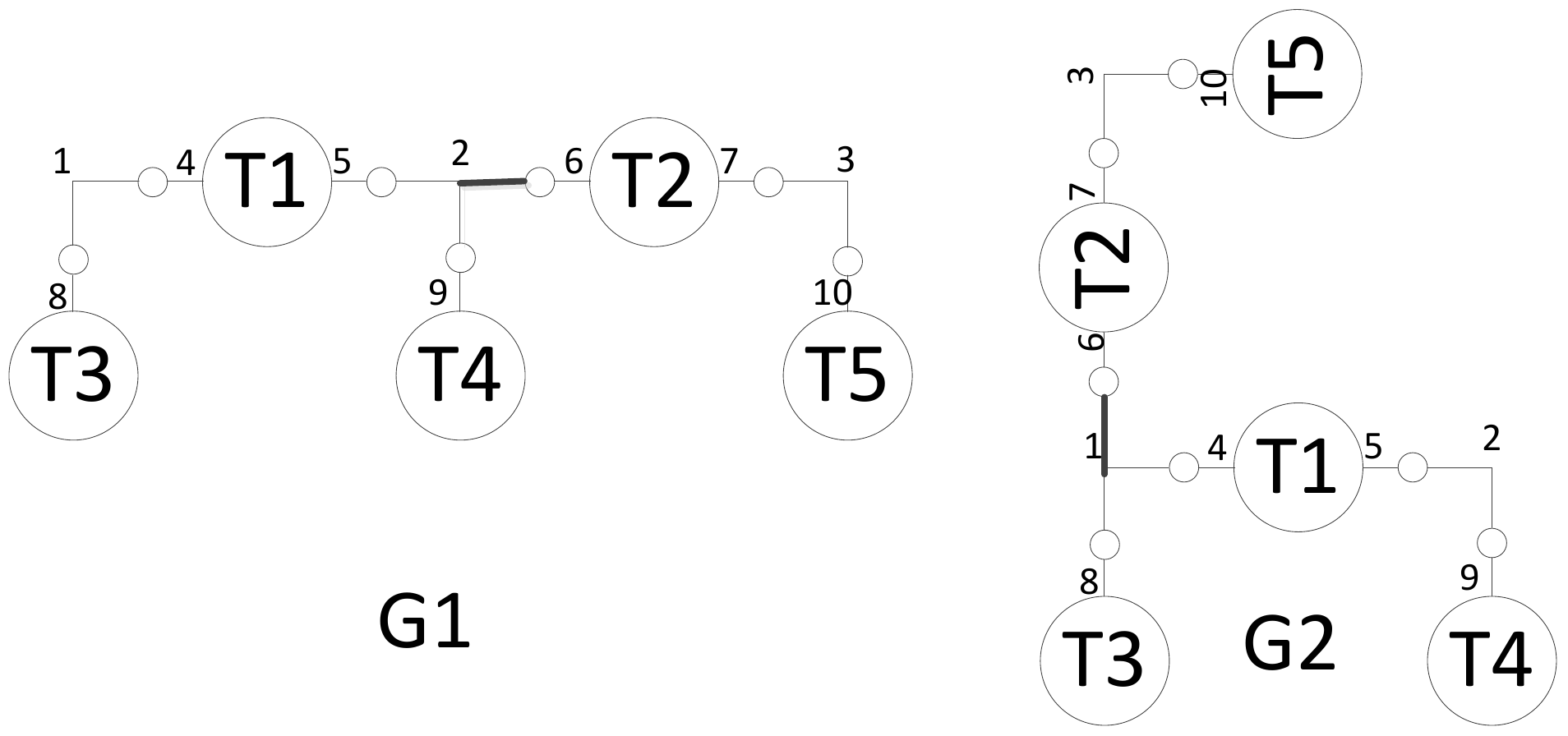}\\
    \caption{The general grafting operation of $G_1\leftrightarrow G_2$}\label{20}
    \end{figure}

    We can draw the general grafting operation of $G_1\leftrightarrow G_2$ at first, as Fig. \ref{20}. It can cover all the grafting operation with anchor nodes $1,2,3$. $G_2$ can be obtained from $G_1$ by cutting the bold edge in the right side of node $2$ and pasting all the subtree of the right side to node $1$.
    In this figure, circles represent  arbitrarily subtrees in $G_1$ and $G_2$. The big circles represent five different location relative to the anchor nodes. The second grafting operation will happen in these circles, where nodes $4-10$ are anchor node candidates of the second grafting operation. And the small circles indicate that nodes $4-10$ are not the unique nodes close to nodes $1,2,3$, but can be any possible nodes at that corresponding position.

\begin{figure}
    \centering
    \includegraphics[width=8cm]{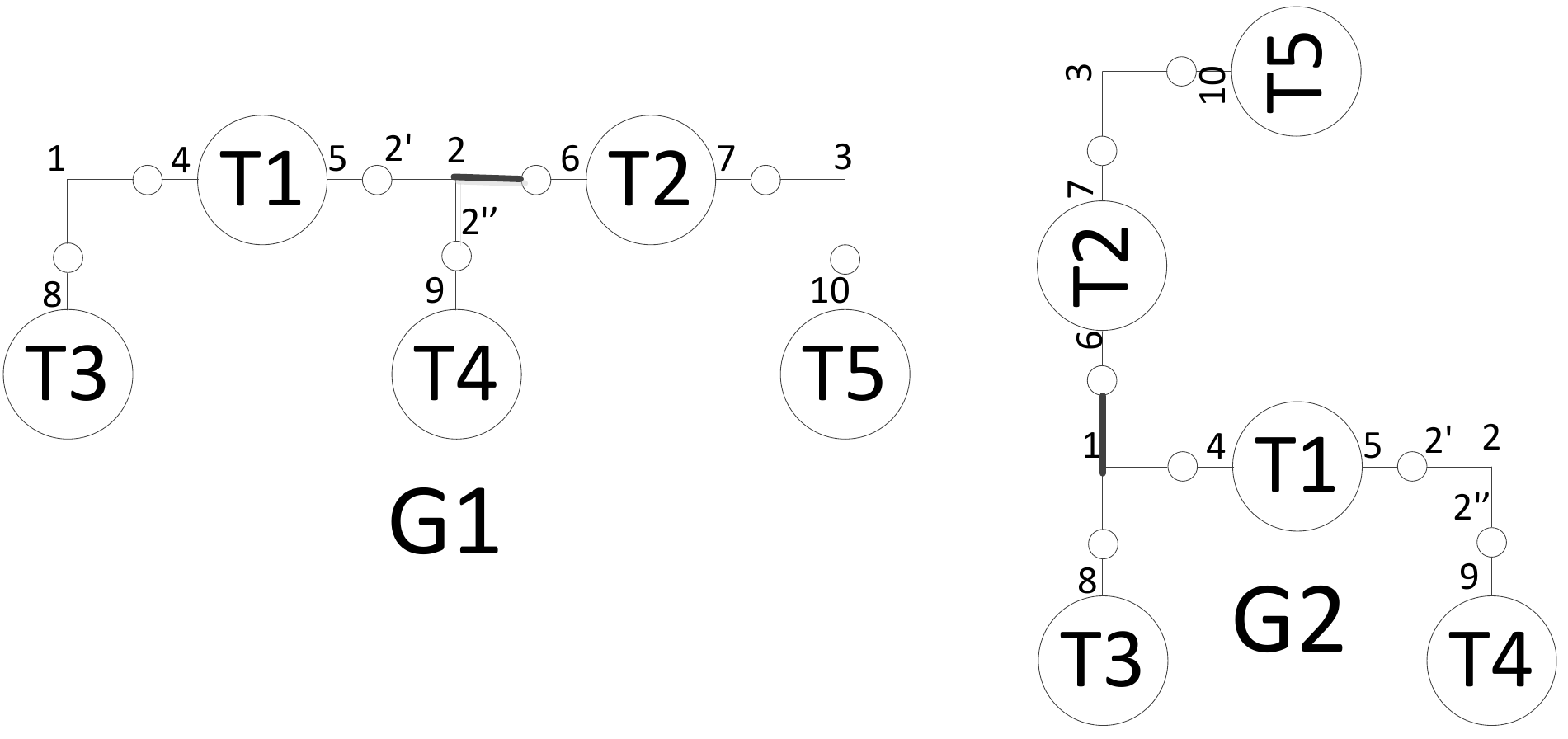}\\
    \caption{Example: the second operation is related with node $2$}\label{21}
    \end{figure}

    Then we will list all the possible grafting patterns of $G_2\leftrightarrow G_3$. We only consider the cases where we cut a subtree outside $T1-T5$ or inside $T1-T5$ and paste it to another position in $T1-T5$. If the operation occurs in small circles, we can expand the big circle to contain the operation.
    If one of nodes $1-3$ be the anchor node of the second grafting operation (node $2$ as an example without loss of generality), we can redraw Fig. \ref{20} into Fig.  \ref{21}. Here we add two auxiliary nodes $2', 2''$ into the figure. The covariance among node $2,2',2''$ approaches to $1$, which means that we can treat them as the same nodes. Then the operation related to node $2$ can be treated in small circles. The only difference between this case and the original one is that some edges in the backbone have to tend to $1$ in this case.

     We only consider the second grafting operation related to the five subtrees, namely, $T_1,T_2,T_3,T_4,T_5$. Among these subtrees, $T_1,T_2$ are in the main path of the anchor nodes $1,2,3$ of the first grafting operation. And $T_3,T_4,T_5$ are out of the main path. So we divide the five subtrees into two sets and divide the second operation to be five different styles, namely 1) inside $T_3,T_4,T_5$ 2) inside $T_1,T_2$ 3) between $T_1,T_2$ 4) among $T_3,T_4,T_5$ 5)between \{$T_1,T_2$\} and \{$T_3,T_4,T_5$\}.

\begin{figure}
    \centering
    \includegraphics[width=8cm]{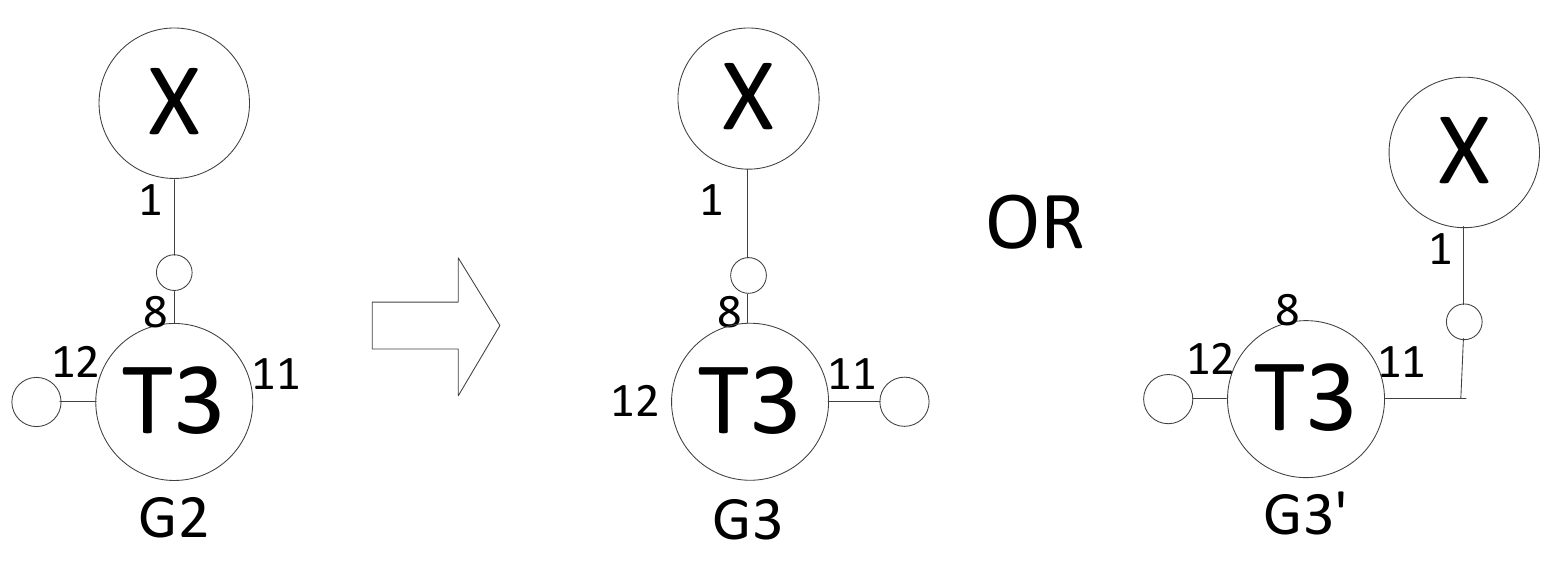}\\
    \caption{Two kinds of cut inside $T_3$}\label{11}
    \end{figure}

    If we want to cut a subtree from $T_i$, there are two methods. We can cut a small subtree inside $T_i$. Otherwise, we can cut all the big tree outside $T_i$ from one side. Taking $T_3$ as an example, the two kinds of cutting are shown in Fig. \ref{11}. For $T_3,T_4,T_5$, the second kind of cut only has one cut point. But it has two different cut points for $T_1$ and $T_2$, because $T_1$ and $T_2$ are in the path of the tree and divide the whole tree into two parts.

\begin{enumerate}[Style 1.]

\item{Grafting operation inside $T_3$, $T_4$ or $T_5$ themselves}

     There are two possible different grafting operations, one for each kind of cut, in these $T_i$. So this subsection has six patterns in total.

\item{Grafting operation inside $T_1$ or $T_2$ themselves}

 For the first kind of cut, there are two different grafting operations in $T_i$, as shown in Fig. \ref{14}. And there are two different grafting patterns for second kind of cut in $T_i$. So this subsection has eight patterns in total.

 \begin{figure}
    \centering
    \includegraphics[width=8cm]{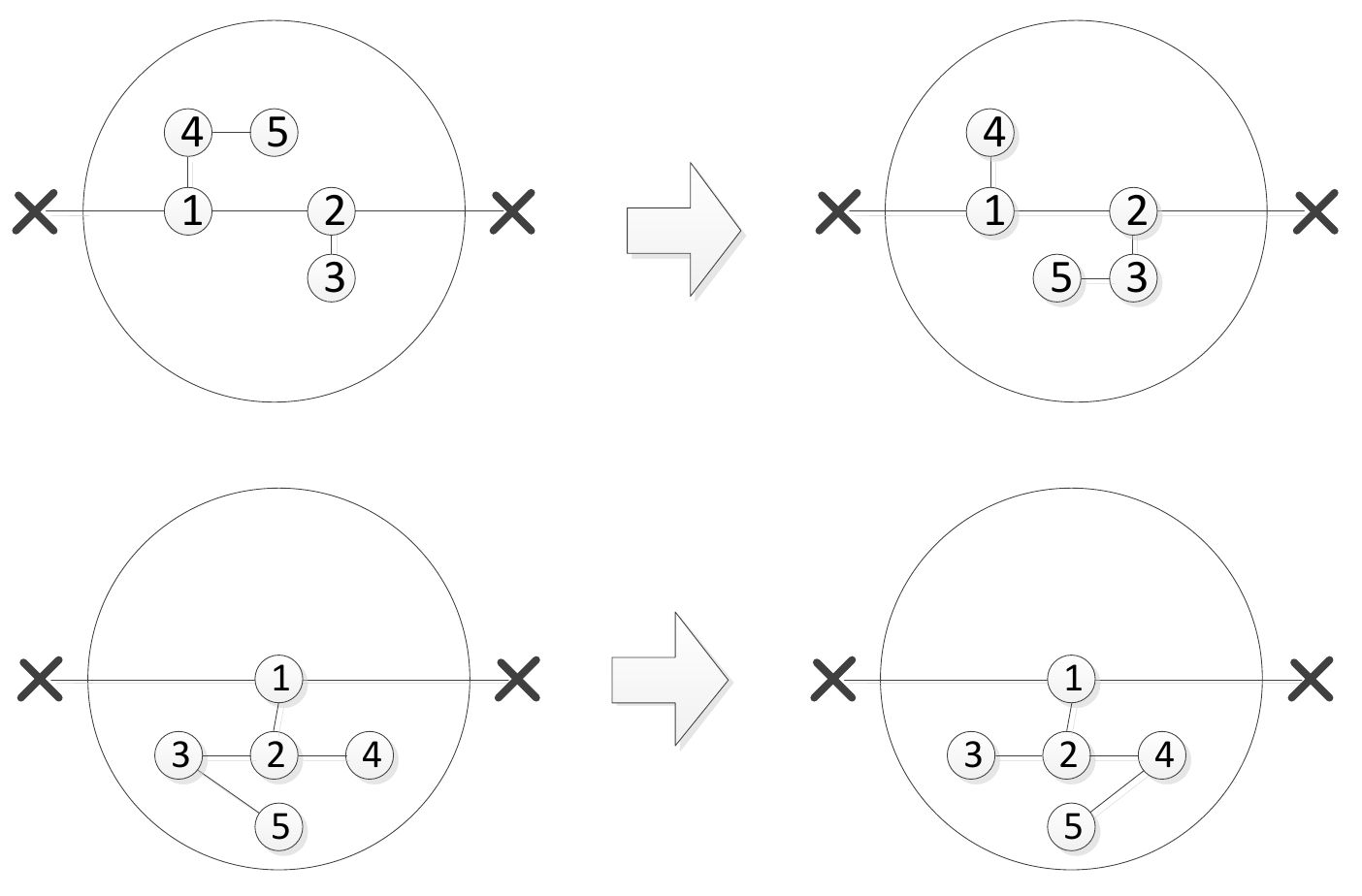}\\
    \caption{The first kind of cut in $T_1$ or $T_2$}\label{14}
    \end{figure}

\item{Grafting operation between $T_1$ and $T_2$}

 For different kind of cut, there is only one pattern of grafting operation from $T_1$ to $T_2$, so as $T_2\rightarrow T_1$. So this subsection has four different patterns in total.

\item{Grafting operation among $T_3$, $T_4$ and $T_5$}

    We only have the first kind of cut in this case for each directed pairs. So this subsection has six different patterns in total.

\item{Grafting operation between \{$T_3$, $T_4$, $T_5$\} and \{$T_1$, $T_2$\}}

For each couple, $T_3-T_1$ as an example, there are three different patterns of grafting operation, two for the first kind in different direction and one for the second kind from $T_1$ to $T_3$. So this subsection has $6\times3=18$ different patterns in total.
\end{enumerate}
We can merge the same patterns and get the classification of Fig. \ref{12} at shown before.

\section{Proof of proposition~\ref{prop7}}\label{A7}

    $\lambda^*$ satisfies $D(\Sigma_{\lambda^*}||\Sigma_2)=D(\Sigma_{\lambda^*}||\Sigma_1)$. When $|\Sigma_1|=|\Sigma_2|$, it will become
    $tr\left(\Sigma_{\lambda^*}( \Sigma_1^{-1}- \Sigma_2^{-1})\right)=0$. Priori knowledge tells us that there is only one single $\lambda^*\in[0,1]$ satisfying $tr\left(\Sigma_{\lambda^*}( \Sigma_1^{-1}- \Sigma_2^{-1})\right)=0$, so we only need to prove $tr\left(\Sigma_{0.5}( \Sigma_1^{-1}- \Sigma_2^{-1})\right)=0$ before we get $\lambda^*=1/2$.

    $tr\left(AB\right)=\sum_{i,j} a_{ij}b_{ij}$ when $A,B$ are both symmetric matrices. Luckily, $A=\Sigma_{0.5}$ and $B=\Sigma_1^{-1}- \Sigma_2^{-1}$ are both symmetric matrices. What's more, $B=\Sigma_1^{-1}- \Sigma_2^{-1}$ only have $2n$ non-zero diagonal elements and $2n$ pairs of non-zero non-diagonal elements, where $n$ is the number of grafting operations. For example, $b_{pp}=\frac{w^2}{1-w^2},b_{qq}=-\frac{w^2}{1-w^2},
    b_{iq}=\frac{w}{1-w^2},b_{ip}=-\frac{w}{1-w^2}$ if we cut node $i$ from node $p$ with edge $e_{ip}$ weighting $w$ and paste it to node $q$, and there aren't any other grafting operations involving these three nodes.
    So $tr\left(\Sigma_{0.5}( \Sigma_1^{-1}- \Sigma_2^{-1})\right)$ is a sum of $4n$ terms, each $4$ terms respecting to one grafting operation. We can divide the polynomial into $n$ parts corresponding to each grafting operation, namely, $tr\left(\Sigma_{0.5}( \Sigma_1^{-1}- \Sigma_2^{-1})\right)=\sum_k P_k$. Then we only need to prove each $P_t$ equals to $0$.

\begin{figure}
  \centering
  \includegraphics[width=8cm]{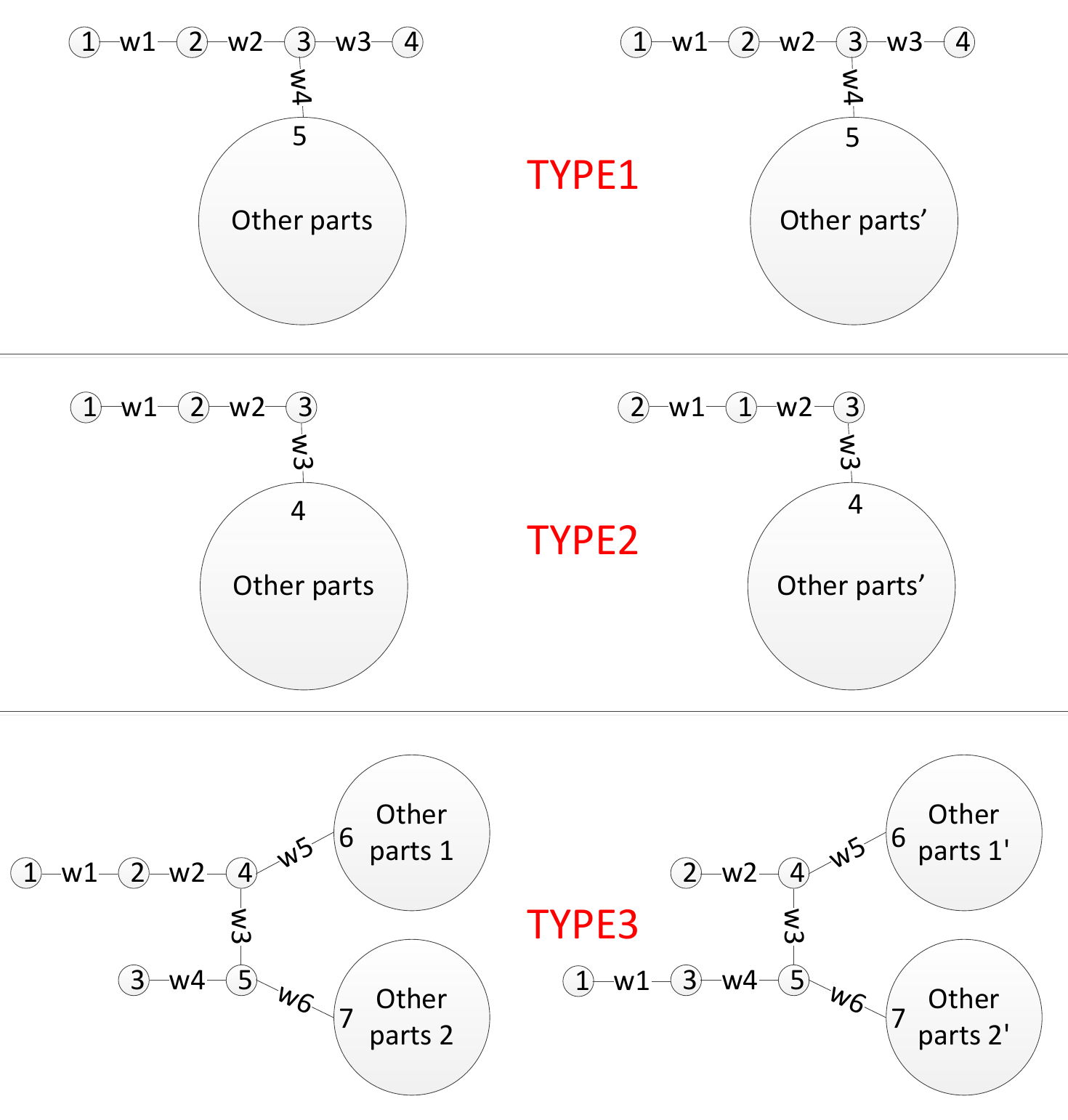}\\
  \caption{Simplified separate grafting operation types}\label{10}
  \end{figure}

    As Fig. \ref{z} shows, there are three different types of grafting operations here.
    After simplification by Proposition \ref{prop4}, the three types of subparts can be translated as shown in Fig. \ref{10}. In this figure, we label these relevant nodes as $1,2,3\dots$ for simplification because it doesn't change the result if we exchange labels of nodes.

    Before we deal with $P_k$ in each case, we will show $\Sigma_{\lambda}^{-1}$ is positive definite matrix. $\Sigma_1$ and $\Sigma_2$ are positive definite matrices, so $\Sigma_1^{-1}$ and $\Sigma_2^{-1}$ are positive definite matrices, namely , $z^T\Sigma_1^{-1} z>0$ and $z^T\Sigma_2^{-1} z>0$ for any non-zero vector $z$. Then $z^T\Sigma_{\lambda}^{-1}z=\lambda z^T\Sigma_1^{-1}z+z^T(1-\lambda)\Sigma_2^{-1}z>0$ for any non-zero vector $z$. So $\Sigma_{\lambda}^{-1}$ is positive definite matrix and its order principal minors are invertible matrices. This result is useful in our later proof.

    We define $f_i=\frac{-w_i}{1-w_i^2}$ and $g_i=\frac{w_i^2}{1-w_i^2}$, so $(1+g_i)g_i=f^2_i$. Then
    we will prove $P_k=0$ in different type of Fig. \ref{10} later.

\subsection{TYPE1}

In this type, we focus on node $1$ to $4$, which is involved in this grafting operation. So

$\Sigma_1^{-1}=$
\[
  \begin{pmat}[{...|}]
  1+g_1&f_1&&&\cr
    f_1&1+g_1+g_2&f_2&&\cr
    &f_2&1+g_2+g_3+g_4&f_3&\mathbf{V} \cr
    &&f_3&1+g_3&\cr\-
    &&\mathbf{V}^T&&\mathbf{K^{(1)}}\cr
  \end{pmat}
\]
$\Sigma_2^{-1}=$
\[
  \begin{pmat}[{...|}]
  1+g_1&&&f_1&\cr
    &1+g_2&f_2&&\cr
    &f_2&1+g_2+g_3+g_4&f_3&\mathbf{V} \cr
    f_1&&f_3&1+g_1+g_3&\cr\-
    &&\mathbf{V}^T&&\mathbf{K^{(2)}}\cr
  \end{pmat}
\]
where $\mathbf{V}$ is a $4\times (N-4)$ matrix with all zero elements except $v_{3,1}=f_4$ and $\mathbf{K^{(1)}},\mathbf{K^{(2)}}$ are covariance matrices of node $5$ to $N$ in $G_1$ and $G_2$ respectively.
And then

$\Sigma_1^{-1}-\Sigma_2^{-1}=$
\[
  \begin{pmat}[{...|}]
  0&f_1&&-f_1&\cr
    f_1&g_1&&&\cr
    &&0&& \cr
    -f_1&&&-g_1&&\cr\-
    &&&&\mathbf{K^{(1)}}-\mathbf{K^{(2)}}\cr
  \end{pmat}
\]
$\Sigma_{0.5}^{-1}=\frac{1}{2}\Sigma_1^{-1}+\frac{1}{2}\Sigma_2^{-1}=$
\[
  \begin{pmat}[{...|}]
  1+\frac{1}{2}g_1&\frac{1}{2}f_1&&\frac{1}{2}f_1&\cr
    \frac{1}{2}f_1&1+g_2+\frac{1}{2}g_1&f_2&&\cr
    &f_2&1+g_2+g_3+g_4&f_3&\mathbf{V}\cr
    \frac{1}{2}f_1&&f_3&1+\frac{1}{2}g_1+g_3&\cr\-
    &\mathbf{V}^T&&&\mathbf{K^{(3)}}\cr
  \end{pmat}
\]
where $\mathbf{K^{(3)}}=\frac{1}{2}\mathbf{K^{(1)}}+\frac{1}{2}\mathbf{K^{(2)}}$
     is an invertible matrix because $\Sigma_{\lambda}^{-1}$ is positive definite matrix. And we define
    \begin{align}
    V{\mathbf{K^{(3)}}}^{-1}V^T=\begin{bmatrix}
    0&&&\\&0&&\\&&\frac{X}{|\mathbf{K^{(3)}}|}&\\&&&0
    \end{bmatrix}\end{align}

    So $P_k$ related to this grafting operation is $g_1m_{22}-g_1m_{44}+2f_1m_{12}-2f_1m_{14}$ where $m_{ij}$ is the $ij$-th element of $\Sigma_{0.5}$.

    $M_{ij}$ is the $(i,j)$ minor of $\Sigma_{0.5}^{-1}$, so
    \begin{align}
    M_{22}=&|\mathbf{K^{(3)}}|(1+g_1)\{(1+g_3)(1+g_2+g_4+\frac{1}{4}g_1)+\frac{1}{4}g_1(g_2+g_4)\}\nonumber\\-&(1+g_1)(1+g_3+\frac{1}{4}g_1)X\nonumber\\
    M_{44}=&|\mathbf{K^{(3)}}|(1+g_1)\{(1+g_2)(1+g_3+g_4+\frac{1}{4}g_1)+\frac{1}{4}g_1(g_3+g_4)\}\nonumber\\-&(1+g_1)(1+g_2+\frac{1}{4}g_1)X\nonumber\\
    M_{12}=&|\mathbf{K^{(3)}}|\{\frac{1}{2}f_1\{(1+g_3)(1+g_2+g_4+\frac{1}{2}g_1)\nonumber\\+&
    \frac{1}{2}g_1(g_2+g_4)\}+\frac{1}{2}f_1f_2f_3\}
    -\frac{1}{2}f_1(1+g_3+\frac{1}{2}g_1)X\nonumber\\
    M_{14}=&|\mathbf{K^{(3)}}|\{\frac{1}{2}f_1\{(1+g_2)(1+g_3+g_4+\frac{1}{2}g_1)\nonumber\\+&
    \frac{1}{2}g_1(g_3+g_4)\}+\frac{1}{2}f_1f_2f_3\}
    -\frac{1}{2}f_1(1+g_2+\frac{1}{2}g_1)X\nonumber
    \end{align}
    and
    \begin{align}
    &2f_1(-M_{12}+M_{14})+g_1(M_{22}-M_{44})=0\\
    P_k=&g_1m_{22}-g_1m_{44}+2f_1m_{12}-2f_1m_{14}\nonumber\\=&
    \left(2f_1(-M_{12}+M_{14})+g_1(M_{22}-M_{44})\right)|\Sigma_{0.5}|=0
    \end{align}

\subsection{TYPE2}

In this type, we focus on node $1$ to $3$, which is involved in this grafting operation. So

$\Sigma_1^{-1}=$
\[
  \begin{pmat}[{..|}]
  1+g_1&f_1&&\cr
    f_1&1+g_1+g_2&f_2&\mathbf{V} \cr
    &f_2&1+g_2+g_3&\cr\-
    &\mathbf{V}^T&&\mathbf{K^{(1)}}\cr
  \end{pmat}
\]
$\Sigma_2^{-1}=$
\[
  \begin{pmat}[{..|}]
  1+g_1+g_2&f_1&f_2&\cr
    f_1&1+g_1&&\mathbf{V}\cr
    f_2&&1+g_2+g_3&\cr\-
    &\mathbf{V}^T&&\mathbf{K^{(2)}}\cr
  \end{pmat}
\]
where $\mathbf{V}$ is a $3\times (N-3)$ matrix with all zero elements except $v_{3,1}=f_3$ and $\mathbf{K^{(1)}},\mathbf{K^{(2)}}$ are covariance matrices of node $4$ to $N$ in $G_1$ and $G_2$ respectively.
And then we use the same progress and get
\begin{align}
    P_k=&g_2m_{22}-g_2m_{11}+2f_2m_{23}-2f_2m_{13}=0
\end{align}
    where $m_{ij}$ is the $ij$-th element of $\Sigma_{0.5}$.

\subsection{TYPE3}

In this type, we focus on node $1$ to $5$, which is involved in this grafting operation. So

$\Sigma_1^{-1}=$
\[
  \begin{pmat}[{....|}]
  1+g_1&f_1&&&&\cr
    f_1&1+g_1+g_2&&f_2&&\cr
    &&1+g_4&&f_4&\mathbf{V} \cr
    &f_2&&1+g_2+g_3+g_5&f_3&\cr
    &&f_4&f_3&1+g_3+g_4+g_6&\cr\-
    &&\mathbf{V}^T&&&\mathbf{K^{(1)}}\cr
  \end{pmat}
\]
$\Sigma_2^{-1}=$
\[
  \begin{pmat}[{....|}]
  1+g_1&&f_1&&&\cr
    &1+g_2&&f_2&&\cr
    f_1&&1+g_1+g_4&&f_4&\mathbf{V} \cr
    &f_2&&1+g_2+g_3+g_5&f_3&\cr
    &&f_4&f_3&1+g_3+g_4+g_6&\cr\-
    &&\mathbf{V}^T&&&\mathbf{K^{(2)}}\cr
  \end{pmat}
\]
where $\mathbf{V}$ is a $5\times (N-5)$ matrix with all zero elements except $v_{4,1}=f_5,v_{5,2}=f_6$ and $\mathbf{K^{(1)}},\mathbf{K^{(2)}}$ are covariance matrices of node $6$ to $N$ in $G_1$ and $G_2$ respectively.
And then we use the same progress and get
\begin{align}
    P_k=&g_1m_{22}-g_1m_{33}+2f_1m_{12}-2f_1m_{13}=0
\end{align}
    where $m_{ij}$ is the $ij$-th element of $\Sigma_{0.5}$.

    $P_k=0$ in all types of grafting operations if these operations are separate. And $tr\left(\Sigma_{0.5}( \Sigma_1^{-1}- \Sigma_2^{-1})\right)=\sum_k P_k=0$.

    So $\lambda^*=\frac{1}{2}$ if all the grafting operations in the chain are separate.

\section{Proof of proposition~\ref{<}}\label{A9}

    We  prove this proposition by using Proposition \ref{prop4} and \ref{combine-graph} repeatedly.

    For $p\leq i\leq j\leq q$, there are $q-p$ grafting operations in the grafting chain $T_p \leftrightarrow \dots \leftrightarrow T_q$. We divide these grafting operations into two sets: grafting operations from $T_i$ to $T_j$ and other grafting operations, namely, set $1$ and set $2$.

    We can simplify tree pairs $(T_p,T_q)$ and  $(T_i,T_j)$ into $(\hat{T}_p,\hat{T}_q)$ and $(\tilde{T}_i,\tilde{T}_j)$  respectively by Proposition \ref{prop4}. $(\hat{T}_p,\hat{T}_q)$ have all the anchor nodes of all $q-p$ operations and the paths of backbone among these operations. $(\tilde{T}_i,\tilde{T}_j)$ have all the anchor nodes of all set $1$ operations and the paths of backbone among these operations. If all the grafting operations in the chain are independent, $(\hat{T}_p,\hat{T}_q)$ have the same substructure with $(\tilde{T}_i,\tilde{T}_j)$ after dropping the extra nodes. So $CI(\hat{T}_p||\hat{T}_q)\geq (\tilde{T}_i||\tilde{T}_j)$ holds by Proposition \ref{combine-graph}. And then $CI(T_i||T_j)\leq CI(T_p||T_q)$.

\begin{figure}
  \centering
  \includegraphics[width=12cm]{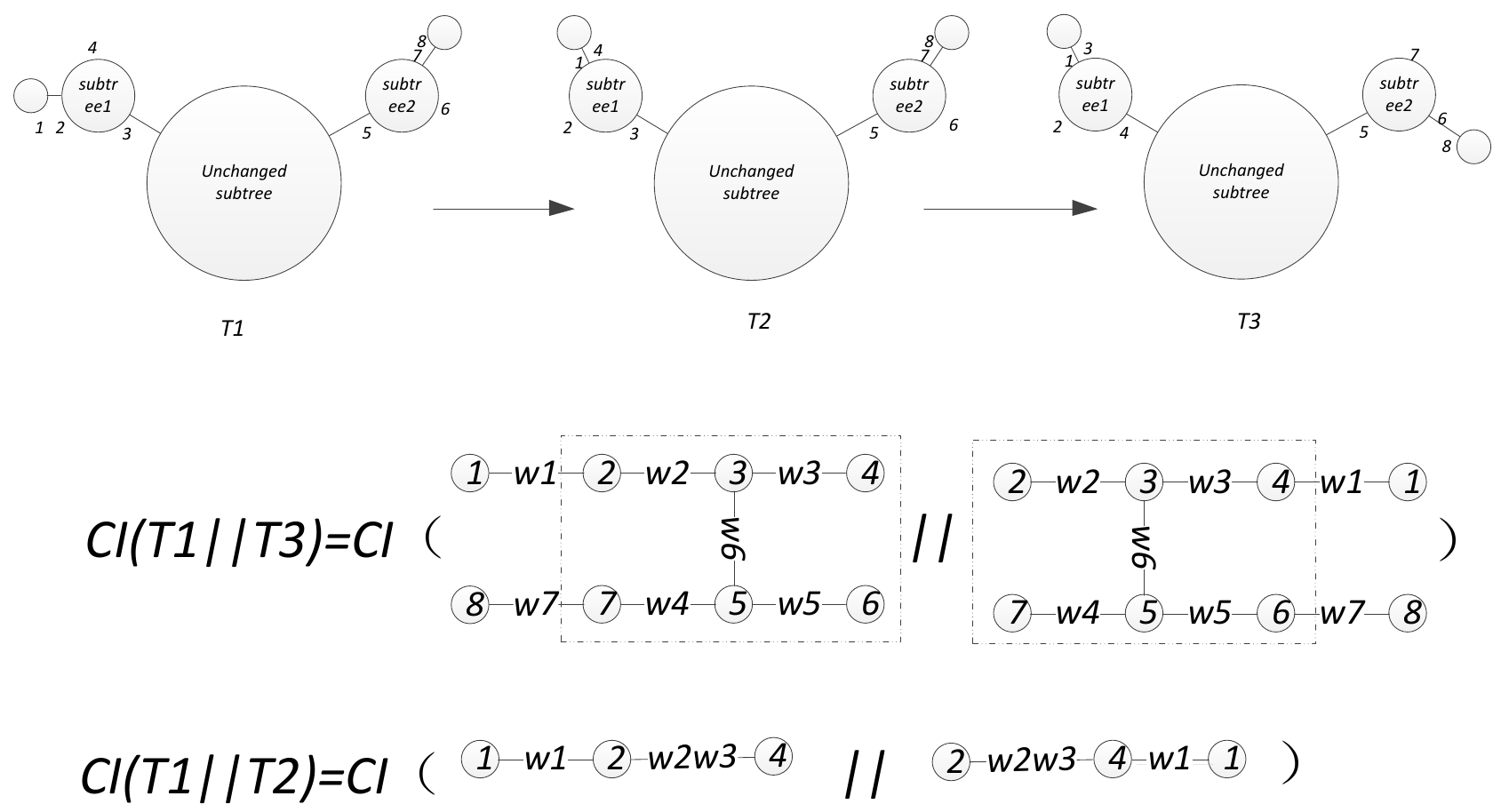}\\
  \caption{Example for Proposition \ref{<}}\label{f<}
\end{figure}

    Here we take a $T_1\leftrightarrow T_2\leftrightarrow T_3$ case as an example, as shown in Fig. \ref{f<}. We can simplify the calculation of $CI(T_1||T_3)$ and $CI(T_1||T_2)$ as shown in Fig. \ref{f<} by Proposition \ref{prop4}. And then Proposition \ref{combine-graph} can tell us $CI(T_1||T_3)\geq CI(T_1||T_2)$. In the same way, we can conclude $CI(T_1||T_3)\geq CI(T_2||T_3)$.

\section{Proof of Proposition~\ref{prop0}}\label{AA1}

Because $\mathbf{\Sigma}_1$ and $\mathbf{\Sigma}_2^{-1}$ are real symmetric positive definite matrices, $\mathbf{\Sigma}_1=\mathbf{L}_1\mathbf{L}_1^T$ and $\mathbf{\Sigma}_2^{-1}=\mathbf{L}_2\mathbf{L}_2^T$ due to Cholesky decomposition where $\mathbf{L}_1$ and $\mathbf{L}_2$ are real non-singular triangular matrices.

The characteristic polynomial of $\mathbf{\Sigma}_1\mathbf{\Sigma}_2^{-1}$ is
\begin{align}
f(\mathbf{\Sigma}_1\mathbf{\Sigma}_2^{-1})&=|\lambda \mathbf{I}-\mathbf{\Sigma}_1\mathbf{\Sigma}_2^{-1}|\nonumber\\&=
|\lambda \mathbf{I}-\mathbf{L}_1\mathbf{L}_1^T\mathbf{L}_2\mathbf{L}_2^T|\nonumber\\&=
|\mathbf{L}_1(\lambda \mathbf{I}-\mathbf{L}_1^T\mathbf{L}_2\mathbf{L}_2^T\mathbf{L}_1)\mathbf{L}_1^{-1}|\nonumber\\&=
|\lambda \mathbf{I}-\mathbf{L}_1^T\mathbf{L}_2\mathbf{L}_2^T\mathbf{L}_1|\nonumber\\&=
|\lambda \mathbf{I}-(\mathbf{L}_1^T\mathbf{L}_2){(\mathbf{L}_1^T\mathbf{L}_2)}^T|
\end{align}
So $\mathbf{\Sigma}_1\mathbf{\Sigma}_2^{-1}$ has the same eigenvalues with $(\mathbf{L}_1^T\mathbf{L}_2){(\mathbf{L}_1^T\mathbf{L}_2)}^T$, which is a real symmetric positive definite matrix. So the eigenvalues of $\mathbf{\Sigma}_1\mathbf{\Sigma}_2^{-1}$ are all positive and $\mathbf{\Sigma}_1\mathbf{\Sigma}_2^{-1}$ is a positive definite matrix.

The eigenvalue decomposition of $\mathbf{\Sigma}_1\mathbf{\Sigma}_2^{-1}$ is $\mathbf{Q}\mathbf{\Lambda} \mathbf{Q}^{-1}$, where $\mathbf{Q}$ is an $N\times N$ matrix and $\mathbf{\Lambda}=Diag(\{\lambda_i\})$ is a diagonal matrix of eigenvalues, in which we put multiple eigenvalues adjacent. $\{\lambda_i\}$ are the eigenvalues of $\mathbf{\Sigma}_1\mathbf{\Sigma}_2^{-1}$, namely, the generalized eigenvalues of $\mathbf{\Sigma}_1$ and $\mathbf{\Sigma}_2$.

And $\mathbf{\Sigma}_1^{(1)}=\mathbf{Q}^{-1}\mathbf{\Sigma}_1{\mathbf{Q}^{-1}}^T=[a_{ij}]$ and $\mathbf{\Sigma}_2^{(1)}=\mathbf{Q}^{-1}\mathbf{\Sigma}_2{\mathbf{Q}^{-1}}^T=[b_{ij}]$ satisfy
$\mathbf{\Sigma}_1^{(1)}=\mathbf{\Lambda} \mathbf{\Sigma}_2^{(1)}$.
\begin{align}
a_{ij}=\lambda_ib_{ij}\\
a_{ij}=a_{ji}=\lambda_jb_{ji}=\lambda_jb_{ij}
\end{align}
So for $\forall i\neq j$, $a_{ij}=b_{ij}=0$ or $\lambda_i=\lambda_j$.

If the eigenvalues of $\mathbf{\Sigma}_1\mathbf{\Sigma}_2^{-1}$ have $k$ different multiple eigenvalues $\lambda^{(1)},\lambda^{(2)},\dots,\lambda^{(k)}$ with multiplicity $n_1,n_2,\dots,n_k$, so
\begin{align}
\mathbf{\Sigma}_2^{(1)}=\begin{bmatrix}
\mathbf{J}_1\\
&\mathbf{J}_2\\
&&\mathbf{J}_3\\
&&&\ddots\\
&&&&\mathbf{J}_k
\end{bmatrix}\\
\mathbf{\Sigma}_1^{(1)}=\begin{bmatrix}
\lambda^{(1)}\mathbf{J}_1\\
&\lambda^{(2)}\mathbf{J}_2\\
&&\lambda^{(3)}\mathbf{J}_3\\
&&&\ddots\\
&&&&\lambda^{(k)}\mathbf{J}_k
\end{bmatrix}
\end{align}
where $\mathbf{J}_i$ is $n_i\times n_i$ inverse matrix.

We can also get an inverse matrix $\mathbf{Q}^{(1)}={\mathbf{\Sigma}_2^{(1)}}^{-0.5}$ so that
\begin{align}
\mathbf{\Sigma}_2^{(2)}=\mathbf{Q}^{(1)}\mathbf{\Sigma}_2^{(1)}{\mathbf{Q}^{(1)}}^T=\mathbf{I}_N\\
\mathbf{\Sigma}_1^{(2)}=\mathbf{Q}^{(1)}\mathbf{\Sigma}_1^{(1)}{\mathbf{Q}^{(1)}}^T=\mathbf{\Lambda}
\end{align}

So we can construct a linear transformation matrix
$\mathbf{P}=\mathbf{Q}^{(1)}\mathbf{Q}^{-1}={\left(\mathbf{Q}^{-1}\mathbf{\Sigma}_2{(\mathbf{Q}^{-1})}^T\right)}^{-\frac{1}{2}}\mathbf{Q}^{-1}$ and thus
\begin{align}
\mathbf{\Sigma}_2'=&\mathbf{P}\mathbf{\Sigma}_2\mathbf{P}^T=\mathbf{I}_N\\
\mathbf{\Sigma}_1'=&\mathbf{P}\mathbf{\Sigma}_1\mathbf{P}^T=\mathbf{\Lambda}
\end{align}

\section{Proof of Proposition~\ref{DR}}\label{AA7}

We want to prove $CI(\hat{\mathbf{\Sigma}}_1||\hat{\mathbf{\Sigma}}_2)\geq CI(\tilde{\mathbf{\Sigma}}_1||\tilde{\mathbf{\Sigma}}_2)$ where $\hat{\mathbf{\Sigma}}_i=\mathbf{A}^*\mathbf{\Sigma}_i{\mathbf{A}^*}^T,
\tilde{\mathbf{\Sigma}}_i=\mathbf{D}\mathbf{\Sigma}_i\mathbf{D}^T$ for
arbitrary $N_O\times N$ matrix $\mathbf{D}$ and $i=1,2$.
Due to equation (\ref{CI2}), we only need to prove $D(\hat{\mathbf{\Sigma}}_\lambda||\hat{\mathbf{\Sigma}}_2)\geq D(\tilde{\mathbf{\Sigma}}_\lambda||\tilde{\mathbf{\Sigma}}_2)$ and $D(\hat{\mathbf{\Sigma}}_\lambda||\hat{\mathbf{\Sigma}}_1)\geq D(\tilde{\mathbf{\Sigma}}_\lambda||\tilde{\mathbf{\Sigma}}_1)$ where
$\hat{\mathbf{\Sigma}}_\lambda^{-1}=\lambda\hat{\mathbf{\Sigma}}_1^{-1}+(1-\lambda)\hat{\mathbf{\Sigma}}_2^{-1}$
and $\tilde{\mathbf{\Sigma}}_\lambda^{-1}=\lambda\tilde{\mathbf{\Sigma}}_1^{-1}+(1-\lambda)\tilde{\mathbf{\Sigma}}_2^{-1}$.

In the following part, I will prove $D(\hat{\mathbf{\Sigma}}_\lambda||\hat{\mathbf{\Sigma}}_1)\geq D(\tilde{\mathbf{\Sigma}}_\lambda||\tilde{\mathbf{\Sigma}}_1)$ and the proof of $D(\hat{\mathbf{\Sigma}}_\lambda||\hat{\mathbf{\Sigma}}_2)\geq D(\tilde{\mathbf{\Sigma}}_\lambda||\tilde{\mathbf{\Sigma}}_2)$ is similar.
\begin{align}
D(\mathbf{\Sigma}_\lambda||\mathbf{\Sigma}_1)=\frac{1}{2}\sum_i g(\nu_i)
\end{align}
where $g(\nu_i)=\ln\left(\lambda+(1-\lambda)\nu_i\right)+\frac{1}{\lambda+(1-\lambda)\nu_i}-1$ and $\{\nu_i\}$ are generalized eigenvalues of $\mathbf{\Sigma}_1$ and $\mathbf{\Sigma}_2$.

$g(1)=0$. $g(\nu_i)$ is decreasing in $(0,1)$ and increasing in $(1,+\infty)$.
If $\nu_1^{(1)}\leq\nu_1^{(2)}\leq\nu_2^{(1)}\leq\nu_2^{(2)}\leq\dots\leq\nu_{N-1}^{(1)}\leq\nu_{N-1}^{(2)}\leq\nu_{N}^{(1)}$, we can choose $\mu_i^{(1)}=\nu_{i+1}^{(1)}$ for $\nu_i^{(2)}\geq 1$ and $\mu_i^{(1)}=\nu_{i}^{(1)}$ for $\nu_i^{(2)}< 1$ so that $\{\mu_i^{(1)}\}$ is an $(N-1)$ subset of $\{\nu_i^{(1)}\}$ and $\frac{1}{2}\sum_{i=1}^{N-1} g(\mu_i^{(1)})\geq\frac{1}{2}\sum_{i=1}^{N-1} g(\nu_i^{(2)})$.

If $m$ elements of $\{\nu_i^{(2)}\}$ are greater than one, $\{\mu_i^{(1)}\}$ contains the first $k$ elements and last $N-1-k$ elements of $\{\nu_i^{(1)}\}$.

We first introduce a preparation of proof.

\subsection{The relationship of eigenvalues}
We have four covariance matrices $\mathbf{\Sigma}_1,\mathbf{\Sigma}_2,\mathbf{\Sigma}_1'=\begin{bmatrix}\mathbf{\Sigma}_1&\mathbf{m}\\\mathbf{m}^T&a_0\end{bmatrix}$ and $\mathbf{\Sigma}_2'=\begin{bmatrix}\mathbf{\Sigma}_2&\mathbf{p}\\\mathbf{p}^T&b_0\end{bmatrix}$.

The eigenvalues of $\mathbf{\Sigma}_1\mathbf{\Sigma}_2^{-1}$ are $\mathbf{\lambda}_1^{(2)}\leq\mathbf{\lambda}_2^{(2)}\leq\dots\leq\lambda_{N-1}^{(2)}$.

The eigenvalues of $\mathbf{\Sigma}_1'\mathbf{\Sigma}_2'^{-1}$ are $\lambda_1^{(1)}\leq\lambda_2^{(1)}\leq\dots\leq\lambda_N^{(1)}$.

For $\mathbf{\Sigma}_1$ and $\mathbf{\Sigma}_2$, we can find an inverse matrix $\mathbf{P}$ where $\mathbf{\Sigma}_1^{(1)}=\mathbf{P}\mathbf{\Sigma}_1\mathbf{P}^T=Diag\{\lambda_1^{(2)},\lambda_2^{(2)},\dots,\lambda_{N-1}^{(2)}\}$ and $\mathbf{\Sigma}_2^{(1)}=\mathbf{P}\mathbf{\Sigma}_2\mathbf{P}^T=\mathbf{I}$. So $\mathbf{P}'=\begin{bmatrix} \mathbf{P}&\\&1\end{bmatrix}$ satisfies $\mathbf{\Sigma}_1'^{(1)}=\mathbf{P}'\mathbf{\Sigma}_1'\mathbf{P}'^T=\begin{bmatrix} \mathbf{\Sigma}_1^{(1)}&\mathbf{a}\\ \mathbf{a}^T&a_0\end{bmatrix}$ and $\mathbf{\Sigma}_2'^{(1)}=\mathbf{P}'\mathbf{\Sigma}_2'\mathbf{P}'^T=\begin{bmatrix} \mathbf{I}&\mathbf{b}\\ \mathbf{b}^T&b_0\end{bmatrix}$, where $\mathbf{a}=\{a_1,a_2,\dots,a_{N-1}\}^T$ and $\mathbf{b}=\{b_1,b_2,\dots,b_{N-1}\}^T$.

So $\{\lambda_k^{(1)}\}$ are the roots of $|\lambda\mathbf{\Sigma}_2'-\mathbf{\Sigma}_1'|=0$, which is equal to $F(\lambda)=0$ where
\begin{align}
F(\lambda)&=|\lambda\mathbf{\Sigma}_2'^{(1)}-\mathbf{\Sigma}_1'^{(1)}|=\left|\begin{matrix} \lambda \mathbf{I}-\mathbf{\Sigma}_1^{(1)}&\lambda\mathbf{b}-\mathbf{a}\\ \lambda\mathbf{b}^T-\mathbf{a}^T&\lambda b_0- a_0\end{matrix}\right|
\nonumber\\&=
\prod_{i=1}^{N-1}(\lambda-\lambda_i^{(2)})\times\nonumber\\&\left\{\lambda b_0- a_0-
(\lambda\mathbf{b}^T-\mathbf{a}^T){(\lambda \mathbf{I}-\mathbf{\Sigma}_1^{(1)})}^{-1}(\lambda\mathbf{b}-\mathbf{a})\right\}
\nonumber\\&=
\prod_{i=1}^{N-1}(\lambda-\lambda_i^{(2)})\times\left\{f_0(\lambda)-\sum_{i=1}^{N-1}\frac{{f_i^2(\lambda)}}{\lambda-\lambda_i^{(2)}}\right\}
\\&=|\mathbf{\Sigma}_2'^{(1)}|\lambda^N+\sum_{i=0}^{N-1} c_i\lambda^i
\end{align}
and $f_i(\lambda)=b_i\lambda-a_i$.

{At first, we assume that there is no multiple
eigenvalues for $\{\lambda_i^{(2)}\}$ and $f_i(\lambda_i^{(2)})\neq0$ for all $1\leq i\leq N-1$}.

 So \begin{align}{(-1)}^{N-k}F(\lambda_{k}^{(2)})=&{(-1)}^{N-k+1}f_k^2(\lambda_{k}^{(2)})\prod_{i=1,i\neq k}^{N-1}(\lambda_{k}^{(2)}-\lambda_i^{(2)})\nonumber\\>&0
 \end{align}
 And $F(+\infty)=+\infty$, $F(-\infty)={(-1)^N}\infty$
 There is at least one root of $F(\lambda)=0$ in $(-\infty,\lambda_1^{(2)})$, $(\lambda_{N-1}^{(2)},+\infty)$ and $(\lambda_{i}^{(2)},\lambda_{i+1}^{(2)})$ for $1\leq i\leq N-2$.

 So we can conclude that $\lambda_1^{(1)}<\lambda_1^{(2)}<\lambda_2^{(1)}<\lambda_2^{(2)}<\dots<\lambda_{N-1}^{(1)}<\lambda_{N-1}^{(2)}<\lambda_{N}^{(1)}$.

{If $\lambda_i^{(2)}$ has multiplicity $n_i$, we can conclude that $\lambda_i^{(2)}$ has multiplicity $n_i-1$ in $F(\lambda)=0$. If we put this exact multiple eigenvalues aside, other eigenvalues satisfy the ordering before.}

{If $f_i(\lambda_i^{(2)})=0$, we can conclude that $\lambda_i^{(2)}$ is also the root of $F(\lambda)=0$.  If we put this eigenvalue aside, other eigenvalues satisfy the ordering before.}

So for all cases, $\lambda_1^{(1)}\leq\lambda_1^{(2)}\leq\lambda_2^{(1)}\leq\lambda_2^{(2)}\leq\dots\leq\lambda_{N-1}^{(1)}\leq\lambda_{N-1}^{(2)}\leq\lambda_{N}^{(1)}$.

\subsection{Main proof}

For $N$-dimension graphs whose covariance matrices are $\mathbf{A}$ and $\mathbf{B}$, we can do inverse linear transformation by $\mathbf{A}'=\mathbf{K}\mathbf{A}\mathbf{K}^T$ and $\mathbf{B}'=\mathbf{K}\mathbf{B}\mathbf{K}^T$. Then we can do dimension-reduction from $N$ to $N-1$ by $\mathbf{A}''=[\mathbf{I}_{N-1},0]\mathbf{A}'[\mathbf{I}_{N-1},0]^T$ and $\mathbf{B}''=[\mathbf{I}_{N-1},0]\mathbf{B}'[\mathbf{I}_{N-1},0]^T$.

The generalized eigenvalues of $\mathbf{A}$ and $\mathbf{B}$ are $\lambda_1^{(1)}\leq\lambda_2^{(1)}\leq\dots\leq\lambda_N^{(1)}$, which are also generalized eigenvalues of $\mathbf{A}'$ and $\mathbf{B}'$.
And the generalized eigenvalues of $\mathbf{A}''$ and $\mathbf{B}''$ are $\lambda_1^{(2)}\leq\lambda_2^{(2)}\leq\dots\leq\lambda_{N-1}^{(2)}$. So $\lambda_1^{(1)}\leq\lambda_1^{(2)}\leq\lambda_2^{(1)}\leq\lambda_2^{(2)}
\leq\dots\leq\lambda_{N-1}^{(1)}\leq\lambda_{N-1}^{(2)}\leq\lambda_{N}^{(1)}$.

We can also do the procedure to reduce the dimension from $N-1$ to $N-2$ and get $\lambda_1^{(2)}\leq\lambda_1^{(3)}\leq\lambda_2^{(2)}\leq\lambda_2^{(3)}
\leq\dots\leq\lambda_{N-2}^{(2)}\leq\lambda_{N-2}^{(3)}\leq\lambda_{N-1}^{(2)}$.

The generalized eigenvalues after $k$ dimension reduction are $\{\lambda_i^{(k+1)}\}$ of size $N-k$.

In No. $(N-N_O)$ dimension reduction stage, we can always find an $N_O$ subset $\{\mu_i^{(N-N_O)}\}$ of $\{\lambda_i^{(N-N_O)}\}$ so that $\frac{1}{2}\sum_{i=1}^{N_O} g\left( \mu_i^{(N-N_O)}\right)\geq \frac{1}{2}\sum_{i=1}^{N_O}g \left( \lambda_i^{(N-N_O+1)}\right)$ because $\lambda_1^{(N-N_O)}\leq\lambda_1^{(N-N_O+1)}\leq\lambda_2^{(N-N_O)}\leq\lambda_2^{(N-N_O+1)}
\leq\dots\leq\lambda_{N_O+1}^{(N-N_O)}\leq\lambda_{N_O}^{(N-N_O+1)}\leq\lambda_{N_O+1}^{(N-N_O)}$.

We can always find $N_O$ subset $\{\mu_i^{(N-N_O-1)}\}$ of $\{\lambda_i^{(N-N_O-1)}\}$ and $\{\mu_i^{(N-N_O)}\}$ of $\{\lambda_i^{(N-N_O)}\}$, which satisfy $\frac{1}{2}\sum_{i=1}^{N_O}g \left( \mu_i^{(N-N_O-1)}\right)\geq \frac{1}{2}\sum_{i=1}^{N_O}g \left( \mu_i^{(N-N_O)}\right)\geq\frac{1}{2}\sum_{i=1}^{N_O} g\left( \lambda_i^{(N-N_O+1)}\right)$ with the same reason.

With the same method, we can conclude that there exist an  $N_O$ subset $\{\mu_i^{(1)}\}$ of $\{\lambda_i^{(1)}\}$ so that $\frac{1}{2}\sum_{i=1}^{N_O} g\left( \mu_i^{(1)}\right)\geq\frac{1}{2}\sum_{i=1}^{N_O} g\left( \lambda_i^{(N-N_O+1)}\right)$. If $k$ elements of $\{\lambda_i^{(N-N_O+1)}\}$ are greater than one, $\{\mu_i^{(1)}\}$ contains the first $k$ elements and last $N_O-k$ elements of $\{\lambda_i^{(1)}\}$.

$\frac{1}{2}\sum_{i=1}^{N_O} g\left( \mu_i^{(1)}\right)$ is $D({\mathbf{\Sigma}}_\lambda||{\mathbf{\Sigma}}_1)$  with linear transformation matrix $\mathbf{A}_k$. \\$\frac{1}{2}\sum_{i=1}^{N_O} g\left( \lambda_i^{(N-N_O+1)}\right)$ is $D(\tilde{\mathbf{\Sigma}}_\lambda||\tilde{\mathbf{\Sigma}}_1)$ under arbitrary linear transformation where the arbitrariness is embodied in $\mathbf{K}$.

So $D({\mathbf{\Sigma}}_\lambda||{\mathbf{\Sigma}}_1)\geq D(\tilde{\mathbf{\Sigma}}_\lambda||\tilde{\mathbf{\Sigma}}_1)$ holds for arbitrary $\mathbf{D}$ and its corresponding $\mathbf{A}_k$.

In the same way, $D({\mathbf{\Sigma}}_\lambda||{\mathbf{\Sigma}}_2)\geq D(\tilde{\mathbf{\Sigma}}_\lambda||\tilde{\mathbf{\Sigma}}_2)$ holds for arbitrary $\mathbf{D}$ and its corresponding $\mathbf{A}_k$.

So $CI(\hat{\mathbf{\Sigma}}_1||\hat{\mathbf{\Sigma}}_2)\geq CI(\tilde{\mathbf{\Sigma}}_1||\tilde{\mathbf{\Sigma}}_2)$ for for arbitrary $\mathbf{D}$ and $\mathbf{A}^*$ is the optimal matrix in the set $\{\mathbf{A}_k|N_O+m-N\leq k\leq m, k\geq0\}$.

\end{document}